\documentclass{lmcs}

\usepackage{lastpage}
\lmcsdoi{16}{4}{9}
\lmcsheading{}{\pageref{LastPage}}{}{}%
{Sep.~25,~2018}{Nov.~10,~2020}{}

\keywords{Lyndon words, ordinals}
\usepackage{amsmath,amssymb,amsthm,amsfonts}
\usepackage{gastex}
\usepackage[plain]{algorithm}    
\usepackage[noend]{algorithmic}  
\usepackage[utf8]{inputenc}
\usepackage{appendix}
\usepackage{hyperref}

\renewcommand{\emptyset}{\varnothing}

\newcommand{\depth}{\operatorname{dp}}
\newcommand{\ltpre}[0]{<_{\text{pre}}}
\newcommand{\lepre}[0]{\leqslant_{\text{pre}}}
\newcommand{\ltlex}[0]{<_{\text{lex}}}
\newcommand{\lelex}[0]{\leqslant_{\text{lex}}}
\newcommand{\gtlex}[0]{>_{\text{lex}}}
\newcommand{\gelex}[0]{\geqslant_{\text{lex}}}
\newcommand{\ltstr}[0]{<_{\text{str}}}
\newcommand{\ltalp}[0]{<_{\text{alp}}}
\newcommand{\gtalp}[0]{>_{\text{alp}}}
\newcommand{\trans}[1]{\mathchoice%
         {\xrightarrow{#1}}
         {\xrightarrow{\smash{\lower1pt\hbox{$\scriptstyle #1$}}}}
         {\text{Erreur}}
         {\text{Erreur}}}

\newcommand{\ropen}[1]{[#1)} 

\begin{document}

\title{Transfinite Lyndon words}

\author[L.~Boasson]{Luc Boasson}
\author[O.~Carton]{Olivier Carton}
\address{IRIF, Université de Paris}
\email{Luc.Boasson@gmail.com, Olivier.Carton@irif.fr}

\maketitle

\begin{abstract}
  In this paper, we extend the notion of Lyndon word to transfinite words.
  We prove two main results.  We first show that, given a transfinite word,
  there exists a unique factorization in Lyndon words that are densely
  non-increasing, a relaxation of the condition used in the case of finite
  words.

  In the annex, we prove that the factorization of a rational word has a
  special form and that it can be computed from a rational expression
  describing the word.
\end{abstract}

\section{Introduction}

Lyndon words were introduced by Lyndon in~\cite{Lyndon54,Lyndon55} as
\emph{standard lexicographic sequences} in the study of the derived series
of the free group over some alphabet~$A$.  These words can be used to
construct a basis of the free Lie algebra over~$A$, and their enumeration
yields Witt's well-known formula for the dimension of the homogeneous
component $\mathcal{L}_n(A)$ of this free Lie algebra.  Lyndon words turn
out to be a powerful tool to prove results such as the ``Runs'' theorem~\cite{Runs17}.  This theorem states that the number of maximal repetitions
in a word of length~$n$ is bounded by~$n$, where a repetition is a factor
which is at least twice as long as its shortest period.

There are several equivalent definitions of these words, but they are usually
defined as those words that are primitive and minimal for the
lexicographic ordering in their conjugacy class.  The nice properties they
enjoy in linear algebra are actually closely related to their properties in
the free monoid.  Lyndon words provide a nice factorization of the free
monoid.

Lyndon words can be studied with the tools of combinatorics on words,
leaving aside the algebraic origin of these words. It then can be proved
directly that each word~$w$ of the free monoid~$A^*$ has a unique
decomposition as a product $w = u_1 \cdots u_n$ of a non-increasing
sequence of Lyndon words $u_1 \gelex \cdots \gelex u_n$ for the
lexicographic ordering.  This uniqueness of the decomposition of each word
is indeed remarkable.  It led Knuth to call Lyndon words \emph{prime words}~\cite[p.~305]{Knuth11}, and we also use this terminology.  As usual, such a
result raises the two following related questions: first, how to
efficiently test whether a given word is prime, and second --- more
ambitious --- how to compute its prime factorization.  It has been shown
that this factorization can be computed in linear time in the size of the
given word~$w$~\cite{Duval80}.

Very often, in the field of combinatorics of words, classical results give
rise to an attempt at some generalization. This can be achieved by adapting
the results to trees or to infinite words. The notion of prime word does
not constitute an exception: unique prime decomposition has already been
extended to $\omega$-words by Siromoney et al.\ in~\cite{Siromoney94}, where
it is shown that any $\omega$-word~$x$ can be uniquely factorized either as
$x = u_0u_1u_2 \cdots $ where ${(u_i)}_{i \ge 0}$ is a non-increasing
sequence of finite prime words, or $x = u_0u_1 \cdots u_n$ where
$u_0,\ldots,u_{n-1}$ is a non-increasing sequence of finite prime words and
$u_n$ is a prime $\omega$-word such that $u_{n-1} \gelex u_n$.  Another
characterization of prime $\omega$-words is provided by~\cite{Melancon96a,Melancon96b} where the prime factorization of some
well-known $\omega$-words such as the Fibonacci word is given.  The prime
factorization of automatic $\omega$-words is still automatic~\cite{GocSaariShallit13}.

The goal of this paper is to extend further such results to transfinite
words, that is, words indexed by countable ordinals.  First we extend the
factorization theorem to all countable words, and second, we provide an
algorithm that computes this factorization for words that can be finitely
described by a rational expression.

The first task is to find a suitable notion of transfinite prime words.
This is not easy, as the different equivalent definitions for finite
prime words do not coincide any more on transfinite words.  Since the
factorization property is presumably their most remarkable one, it can be
used as a gauge to measure the accuracy of a definition.  If a definition
allows us to prove that each transfinite word has a unique decomposition in
prime words, it can be considered as the right one.  The two main points are
that the factorization should always exist and that it should be unique.
Of course, the definition should also satisfy the following additional
requirement: it has to be an extension of the classical one for finite
words, meaning that it must coincide with the classical definition for
finite words.  We introduce such a definition.  The existence and
uniqueness of the factorization is obtained by slightly relaxing the property
of being non-increasing.  It is replaced by the property of being densely
non-increasing (see Section~\ref{sec:factorization} for the precise
definition).  As requested, the two properties coincide for finite
sequences.  Our results extend the ones of Siromoney et
al.~\cite{Siromoney94}, as we get the same definition of prime words of
length~$\omega$ and the same decomposition for words of length~$\omega$.

The second task is to extend the algorithmic property of the decomposition
of a word in prime words.  Of course, it is not possible to compute the
factorization of any transfinite word, but we have focused on the so-called
\emph{rational words}, that is, words that can be described from the
letters using product and $\omega$-operations (possibly nested).  We prove
that the factorization of these rational words have a special form.  It can
be a transfinite sequence of primes, but only finitely many different ones
occur in it.  Furthermore, all the prime words occurring are also rational
and the sequence is really non-increasing in that case.  We give an
algorithm that computes the factorization of a rational word given by an
expression involving products and $\omega$-operations.

The paper is organized as follows.  Basic definitions of ordinals and
transfinite words are recalled in Section~\ref{sec:preliminaries}.  The
definition of prime words is given in Section~\ref{sec:prime}, with a few
properties used in the rest of the paper.  The existence and uniqueness of
the prime factorization is proved in Section~\ref{sec:factorization}.  The
Appendix~\ref{sec:rational} is devoted to rational words and to the
properties of their prime factorization.  The algorithm to compute this
prime factorization is described and proved in Appendix~\ref{sec:algo}.  A
short version of this paper has been published in the proceedings of
DLT'2015~\cite{BoassonCarton2015}.

\section{Preliminaries}\label{sec:preliminaries}

In this section, we first recall the notion of an ordinal and the notion of
a transfinite word, that is, a sequence of letters indexed by an ordinal.

\subsection{Ordinals}

We give in this section a short introduction to ordinals but we assume that
the reader is already familiar with this notion.  We do not define formally
all notions.  We refer the reader to Rosenstein~\cite{Rosenstein82} for a
complete introduction to the theory of ordinals. In this paper, we only use
countable ordinals.  As an abuse of language, we use throughout the paper
the word \emph{ordinal} for \emph{countable ordinal}.  An ordinal is an
isomorphism class of well-founded countable linear (that is total)
orderings.  The symbol~$\omega$ denotes, as usual, (the isomorphism class
of) the ordering of the non-negative integers.  Here we give a few examples
of ordinals.  The ordinal $\omega\cdot2$ is the ordering made of two copies
of~$\omega$: $0,2,4, \ldots, 1,3,5, \ldots$.  More generally, the ordinal
$\omega \cdot k$ is the ordinal made of~$k$ copies of~$\omega$.  The
ordinal~$\omega^2$ is the lexicographic ordering of pairs of non-negative
integers: $(m_2,m_1) < (m'_2,m'_1)$ holds if either $m_1 < m'_1$ holds or
both $m_1 = m'_1$ and $m_2 < m'_2$ hold.  Note that the rightmost
components are compared first.  This ordinal~$\omega^2$ can be seen as
$\omega$ copies of~$\omega$. More generally, the ordinal~$\omega^k$ for a
fixed~$k \ge 0$ is the lexicographic ordering of $k$-tuples
$(m_k,\ldots,m_1)$ of non-negative integers.  The ordering $\omega^\omega$
is the ordering on $k$-tuples, $(m_k,\ldots,m_1)$ for $k$ ranging over all
non-negative integers, defined as follows.  The relation
$(m_k,\ldots,m_1) < (m'_{k'},\ldots,m'_1)$ holds in $\omega^\omega$ if
either $k < k'$ holds or both $k = k'$ and
$(m_k,\ldots,m_1) < (m'_k,\ldots,m'_1)$ holds in~$\omega^k$.

An ordinal~$\alpha$ is said to be a \emph{successor} if
$\alpha = \beta + 1$ for some ordinal~$\beta$.  An ordinal is called
\emph{limit} if it is neither $0$, nor a successor ordinal. As usual, we
identify the linear ordering on ordinals with membership: an
ordinal~$\alpha$ is then identified with the set of ordinals smaller
than~$\alpha$.  In this paper, we mainly use ordinals to index sequences.
Let $\alpha$ be an ordinal.  A sequence~$x$ of length~$\alpha$ (or an
$\alpha$-sequence) of elements from a set~$E$ is a function which maps any
ordinal~$\beta$ smaller than~$\alpha$ to an element of~$E$.  A sequence~$x$
is usually denoted by $x = {(x_\beta)}_{\beta<\alpha}$.  A subset~$\Omega$ of
ordinals is called \emph{closed} if it is closed under taking limit: if
$\alpha = \sup \; \{ \alpha_n \mid n < \omega\}$ with $\alpha_n \in \Omega$
for each~$n < \omega$, then $\alpha \in \Omega$.  Note that any bounded
closed subset of ordinals has a greatest element.  This holds because we
only consider countable ordinals as already mentioned.  Let $\gamma$ and
$\gamma'$ be two ordinals such that $\gamma\le\gamma'$.  There exists a
unique ordinal denoted by $\gamma'-\gamma$ such that
$\gamma + (\gamma'-\gamma) = \gamma'$.  We let $\ropen{\gamma,\gamma'}$ denote
the interval $\{ \beta \mid \gamma \le \beta < \gamma'\}$.  It is empty if
$\gamma' = \gamma$ and it only contains $\gamma$ if $\gamma' = \gamma+1$.
If $\gamma'$ is a successor ordinal, that is, if $\gamma' = \gamma''+1$ for
some ordinal~$\gamma''$, the interval $\ropen{\gamma,\gamma'}$ is also denoted by
$[\gamma,\gamma'']$.

\subsection{Words}\label{sec:words}

Let $A$ be a finite set called the \emph{alphabet} equipped with a linear
ordering~$\ltalp$.  Its elements are called \emph{letters}.  In the
examples, we often assume that $A = \{a,b\}$ with $a \ltalp b$.  This
ordering on~$A$ is necessary to define the lexicographic ordering on words.
For an ordinal~$\alpha$, an $\alpha$-sequence of letters is also called a \emph{word}
of length~$\alpha$ or an \emph{$\alpha$-word} over~$A$.  The sequence of length~$0$
which has no element is called the \emph{empty word} and it is denoted
by~$\varepsilon$.  The length of a word~$x$ is denoted by~$|x|$.  The set of all
words of countable length over~$A$ is denoted by $A^{\#}$.

Let $x$ be a word ${(a_\beta)}_{\beta<\alpha}$ of length~$\alpha$.  For any
$\gamma \le \gamma'\le \alpha$, we let $x\ropen{\gamma,\gamma'}$ denote the word
${b_\beta}_{\beta<\gamma'-\gamma}$ of length $\gamma'-\gamma$ defined by
$b_\beta = a_{\gamma+\beta}$ for any $0 \le \beta < \gamma'-\gamma$.  It is
the empty word if $\gamma' = \gamma$ and it is a single letter if
$\gamma' = \gamma+1$.  Such a word $x\ropen{\gamma,\gamma'}$ is called a
\emph{factor} of~$x$.  A word of the form $x\ropen{0,\gamma}$ (resp.,
$x\ropen{\gamma,\alpha}$) for $0 \le \gamma \le \alpha$ is called a \emph{prefix}
(resp., \emph{suffix}) of~$x$. The prefix (resp., suffix) is called
\emph{proper} whenever $0<\gamma<\alpha$.  If $x$ is the word
${(ab)}^\omega {(bc)}^\omega$ of length~$\omega\cdot2$, the prefix
$x\ropen{0,\omega}$ is ${(ab)}^\omega$, the suffix $x\ropen{\omega,\omega\cdot2}$ is
${(bc)}^\omega$ and the factor $x\ropen{5,\omega+2}$ is the word ${(ba)}^\omega bc$.
Notice that a proper suffix of a word~$x$ may be equal to~$x$.  For
instance, the proper suffix $x\ropen{4,\omega\cdot2}$ of the word
$x = {(ab)}^\omega {(bc)}^\omega$ is equal to~$x$.  Notice however that a
proper prefix~$y$ of a word~$x$ cannot be equal to~$x$, since it satisfies
$|y| < |x|$.

The \emph{concatenation}, also called the \emph{product}, of two words
$x = {(a_\gamma)}_{\gamma<\alpha}$ and $y = {(b_\gamma)}_{\gamma<\beta}$ of
lengths $\alpha$ and~$\beta$ is the word
$z = {(c_\gamma)}_{\gamma<\alpha+\beta}$ of length $\alpha+\beta$ given by
$c_\gamma = a_\gamma$ if $\gamma<\alpha$ and $c_\gamma = b_{\gamma-\alpha}$
if $\alpha \le \gamma < \alpha+\beta$.  This word is merely denoted
by~$xy$.  Note that the product~$xy$ may be equal to~$y$ even if $x$ is
non-empty: take for instance $x = a$ and $y = a^\omega$.  Note that a
word~$x$ is a prefix (resp., suffix) of a word~$x'$ if $x' = xy$ (resp.,
$x' = yx$) for some word~$y$.  The word~$x$ is a factor of a word~$x'$ if
$x' = yxz$ for two words $y$ and~$z$.  Note that for any word~$x$ and for
any ordinals $\gamma \le \gamma' \le \gamma'' \le |x|$, the equality
$x\ropen{\gamma,\gamma''} = x\ropen{\gamma,\gamma'}x\ropen{\gamma',\gamma''}$ holds.

More generally, let ${(x_\beta)}_{\beta<\alpha}$ be an $\alpha$-sequence of
words.  The word obtained by concatenating the words of the sequence
${(x_\beta)}_{\beta<\alpha}$ is denoted by $\prod_{\beta<\alpha}{x_\beta}$.
Its length is the sum $\sum_{\beta<\alpha}{|x_\beta|}$.  The product
$\prod_{n<\omega}{x}$ for a given word~$x$ is denoted by~$x^\omega$.  An
\emph{$\alpha$-factorization} of a word~$x$ is a sequence
${(x_\beta)}_{\beta<\alpha}$ of words such that
$x = \prod_{\beta<\alpha}{x_\beta}$.

We write $x \lepre x'$ whenever $x$ is a prefix of~$x'$ and $x \ltpre x'$
whenever $x$ is a prefix of~$x'$ different from~$x'$. The relation~$\ltpre$
is an ordering on~$A^{\#}$.  The ordering~$\ltstr$ is defined by
$x \ltstr x'$ if there exist two letters $a \ltalp b$ and three words $y$,
$z$ and~$z'$ such that $x = yaz$ and $x' = ybz'$.  The \emph{lexicographic
  ordering}~$\lelex$ is finally defined by $x \lelex x'$ if $x \lepre x'$
or $x \ltstr x'$.  We write $x \ltlex x'$ whenever $x \lelex x'$ and
$x \neq x'$.  The relation~$\ltlex$ is a linear ordering on~$A^{\#}$.  Note
that the ordering~$\ltalp$ is the restriction of~$\ltstr$ to the alphabet.

Let ${(x_n)}_{n<\omega}$ be a sequence of words such that
$x_n \lepre x_{n+1}$ for each $n \ge 0$.  By definition, the \emph{limit}
of the sequence ${(x_n)}_{n<\omega}$ is the product $\prod_{n<\omega}{u_n}$
where $u_0 = x_0$ and each word~$u_{n+1}$ for $n \ge 0$ is the unique word
such $x_{n+1} = x_{n}u_n$.

We mostly use Greek letters $\alpha,\beta,\ldots$ to denote ordinals,
letters $a,b,\ldots$, to denote elements of the alphabet, letters
$x,y,\ldots$ to denote transfinite words and letters $u,v,\ldots$ to denote
prime transfinite words.

\section{Prime words}\label{sec:prime}

In this section, we introduce the crucial definition of a prime transfinite
word.  We also prove some basic properties of these words, as well as some
closure properties.  All these preliminary results are helpful to prove
the existence of the prime factorization.  We start with the classical
definition of a primitive word.

A word~$x$ is \emph{primitive} if it is not the power of another word,
i.e., if the equality $x = y^\alpha$ for some ordinal~$\alpha$ and some word~$y$
implies $\alpha = 1$ and $y = x$.  Note that any word~$x$ is either primitive or
the power~$y^\alpha$ of some primitive word~$y$ for some ordinal $\alpha \ge 2$~\cite{CartonChoffrut01}.

\begin{defi}
  A word~$w$ is prime, also called Lyndon, if $w$ is
  primitive and any proper suffix~$x$ of~$w$ satisfies 
\end{defi}
The terminology \emph{prime} is borrowed from~\cite[p.~305]{Knuth11}.  It
is justified by Theorem~\ref{thm:main}, which states that any word has a
unique factorization in prime words which is almost non-increasing (see
Section~\ref{sec:factorization} for a precise statement).

\begin{exa}
  Both finite words $a^2b$ and $a^2bab$ are prime.  Both finite words $aba$
  and $abab$ are not prime.  Indeed, the suffix~$a$ of $aba$ satisfies
  $a \ltlex aba$ and $abab$ is not primitive.  The $\omega$-words
  $ab^\omega$ and $abab^2ab^3ab^4 \cdots$ are prime.  Both $\omega$-words
  $ba^\omega$ and ${(ab)}^\omega$ are not prime.  Indeed, the
  suffix~$a^\omega$ of~$ba^\omega$ satisfies $a^\omega \ltlex ba^\omega$
  and the $\omega$-word ${(ab)}^\omega$ is not primitive.
\end{exa}

Let us make some comments about this definition.  First note that only
proper suffixes are considered, since the empty word~$\varepsilon$ is a
suffix of any word~$w$ but does not satisfy $w \lelex \varepsilon$ (unless
$w = \varepsilon$).  Second each suffix~$x$ of a prime word~$w$ must
satisfy $w \lelex x$, that is, either $w \lepre x$ or $w \ltstr x$.  Since
the length of~$x$ is smaller than or equal to the length of~$w$, the
relation $w \ltpre x$ is impossible, since $w \ltpre x$ would imply
$|w| < |x|$.  The relation $w \lepre x$ reduces then to $w = x$.
Therefore, a word~$w$ is prime if it is primitive and any proper suffix~$x$
of~$w$ satisfies either $w = x$ or $w \ltstr x$.  This last remark is so
frequently used along the paper that it is not quoted.

Our definition of prime words coincides with the classical definition for
finite words~\cite[Chap.~5]{Lothaire}.  A finite word is a prime word if it
is minimal in its conjugacy class or, equivalently, if it is strictly
smaller than any of its proper suffixes~\cite[Prop.~5.1.2]{Lothaire}.  A
proper suffix of a finite word cannot be equal to the whole word and
therefore, it does not matter whether it is required that any proper suffix
is \emph{strictly smaller} or just \emph{smaller} than the whole word.  For
transfinite words, it does matter, since some proper suffix might be equal
to the whole word.  Our definition indeed allows a suffix of a prime word
to be equal to the whole word.  The word $w = a^\omega b$ of length
$\omega+1$ is prime, but some of its proper suffixes such as
$w\ropen{1,\omega+2}$ or $w\ropen{2,\omega+2}$ are equal to~$w$.

Our definition also requires the word to be primitive.  It is not needed
for finite words, since, in that case, being smaller than all its suffixes
implies primitivity. Indeed, if the finite word~$x$ is equal to~$y^n$ for
$n \ge 2$, then $y$ is a proper suffix of~$x$ that is strictly smaller
than~$x$.  Therefore, $x$ cannot be prime.  This argument no longer holds
for transfinite words.  Of course, the $\omega$-word $x = a^\omega$ is not
primitive, but none of its proper suffixes is strictly smaller than itself.
Each proper suffix of~$x$ is actually equal to~$x$.  The same property
holds for each word of the form $a^\alpha$ where $\alpha$ is a power
of~$\omega$, that is, $\alpha = \omega^\beta$ for some ordinal
$\beta \ge 1$.

Our definition of prime words also coincides with the definition for
$\omega$-words given in~\cite{Siromoney94} where an $\omega$-word is called
prime if it is the limit of finite prime words.  It is also shown
in~\cite[Prop.~2.2]{Siromoney94} that an $\omega$-word is prime if and only
if it is strictly smaller than any of its suffixes.  Requiring that no
suffix is equal to the whole $\omega$-word prevents the $\omega$-word from
being periodic, that is, of the form~$x^\omega$ for some finite word~$x$.
These last words are the only non-primitive $\omega$-words.  Let us now
give a more involved example.

\begin{exa}\label{exa:involved}
  Define the sequence ${(u_n)}_{n < \omega}$ of words inductively by
  $u_0 = a$ and $u_{n+1} = u_n^\omega b$.  The first words of the sequence
  are $u_1 = a^\omega b$ and $u_2 = {a^\omega b}^\omega b$.  It can be
  proved by induction on~$n$ that the length of~$u_n$ is $\omega^n+1$ since
  $(\omega^n+1)\cdot\omega + 1 = \omega^{n+1}+1$.  Let $u_\omega$ be the
  word $u_0u_1u_2 \cdots$ of length $\omega^\omega$.  Note that the
  equality $u_{n}u_{n+1} = u_{n+1}$ holds for each $n \ge 0$ and therefore
  the equality $u_\omega = u_{n}u_{n+1}u_{n+2} \cdots$ also holds for each
  $n \ge 0$.  The word $u_\omega$ is actually the limit of the sequence
  ${(u_n)}_{n<\omega}$ as defined in Section~\ref{sec:words}.  The limit of
  $\omega^n+1$ is $\omega^\omega$, the length of~$u_\omega$ and the prefix
  of length~$\omega^n+1$ of~$u_\omega$ coincides with $u_n$.  It is proved
  later that each word~$u_n$ is prime and that their limit~$u_\omega$ is
  also prime.
\end{exa}

\subsection{Properties of prime words}\label{sec:properties}

The following results are useful tools for proving that a given word~$w$ is
prime.  The next lemma makes it easier to prove that $w$ is primitive when
it has already been shown that $w$ is smaller than each of its suffixes.

\begin{lem}\label{lem:power}
  Let $x$ be a word of the form $y^\alpha$ for some word~$y$ and some
  ordinal~$\alpha$.  If $\alpha$ is not a power of~$\omega$, that is, if
  $\alpha \neq \omega^\beta$ for any $\beta \ge 0$ (with $\omega^0 = 1$),
  there exists a suffix~$z$ of~$x$ such that $z \ltpre x$.  If $\alpha$ is
  equal to~$\omega^\beta$ for some $\beta \ge 1$, then every non-empty
  suffix~$z$ of~$x$ has a suffix equal to~$x$.
\end{lem}
\begin{proof}
  Let $\alpha = \omega^{\beta_1} + \cdots + \omega^{\beta_n}$ be the Cantor
  normal form of~$\alpha$ where $\beta_1 \ge \cdots \ge \beta_n$.  If
  $\alpha$ is not a power of~$\omega$, then $n \ge 2$ and
  $\omega^{\beta_n} < \alpha$.  It follows that the word
  $z = y^{\omega^{\beta_n}}$ is a proper suffix and a proper prefix
  of~$y$. The last statement directly follows from the following property
  of powers of~$\omega$: if $\alpha = \omega^\beta$ and
  $\alpha = \alpha_1 + \alpha_2$, then either $\alpha_2 = 0$ or
  $\alpha_2 = \alpha$.  The former case is excluded because $z$ is
  non-empty and the result is trivial in the latter case.
\end{proof}

\begin{lem}\label{lem:vequalphaupw}
  Let $u$ and $v$ be two prime words such that $u \ltlex v$.  Then $v$ can
  be factorized as $v = u^\alpha xy$ for some ordinal~$\alpha$ and words
  $x$ and~$y$ such that $|x| \le |u|$ and $u \ltstr x$.
\end{lem}
\begin{proof}
  Let $\alpha$ be the greatest ordinal such that $u^\alpha$ is a prefix
  of~$v$.  This ordinal does exist, since the set of ordinals~$\alpha$ such
  that $u^\alpha$ is a prefix of~$v$ is closed.  The word~$v$ is then equal
  to~$u^\alpha z$ for some word~$z$.  Let us define the words $x$ and~$y$
  as follows: if $|z| \le |u|$, let $x = z$ and let $y = \varepsilon$.  If
  $|u| \le |z|$, let $x = z\ropen{0,|u|}$ and let $y = z\ropen{|u|,|z|}$.  Note that
  the two definitions coincide if $|u| = |z|$.  In both cases, the equality
  $z = xy$ holds and $x$ satisfies $|x| \le |u|$.  We claim that
  $u \ltstr x$.  It suffices to prove that $u \ltlex x$ since
  $|x| \le |u|$.  First note that the equality $u = x$ contradicts the
  definition of~$\alpha$ and is therefore impossible.  Suppose, by
  contradiction, that $x \ltlex u$, that is, either $x \ltpre u$ or
  $x \ltstr u$.  The case $x \ltpre u$ only occurs if $|x| < |u|$, that is,
  if $|z| < |u|$.  In that case $x$ is equal to~$z$ and is a suffix of~$v$.
  If $\alpha = 0$, $x$ is also a prefix of~$v$.  If $\alpha > 0$, then $u$
  is a prefix of~$v$ and $x \ltlex u$.  In both cases, this is a
  contradiction since $v$ is prime.  If $x \ltstr u$, the suffix~$xy$
  of~$v$ satisfies $xy \ltlex v$, and this is again a contradiction since
  $v$ is prime.
\end{proof}

Note that the hypothesis of the previous lemma can be weakened.  Indeed,
the only required assumptions are that $u \ltlex v$ and that each proper
suffix of~$v$ is larger than~$v$.

\begin{cor}\label{cor:ualphav}
  Let $u$ and $v$ be two prime words such that $u \ltlex v$.  Then
  $u^\alpha \ltlex u^\alpha v \lelex v$ holds for every ordinal~$\alpha$.
\end{cor}
\begin{proof}
  The first relation $u^\alpha \ltlex u^\alpha v$ is straightforward. By
  Lemma~\ref{lem:vequalphaupw}, the word~$v$ is equal to $u^\beta xy$ for
  some ordinal~$\beta$ and some words $x$ and~$y$ such that $|x| \le |u|$
  and $u \ltstr x$.  The word~$u^\alpha v$ is then equal to
  $u^{\alpha+\beta} xy$.  If $\alpha+\beta = \beta$, then $u^\alpha v = v$.
  If $\alpha+\beta > \beta$, then $u^\alpha v \ltlex v$ since $u \ltstr x$.
\end{proof}

\subsection{Closure properties}

In this section, we prove some results which state that, under some
hypothesis, the product of some words yields a prime word.  To some extend,
these results generalize the classical results on finite words.

It is well-known that if two finite prime words $u$ and~$v$ satisfy $u \ltlex
v$, then the word~$uv$ is prime and satisfies $u \ltlex uv \ltlex v$~\cite[Prop.~5.1.3]{Lothaire}.  It can easily be shown by induction on~$n$
that $u^{n}v$ is also prime for any integer~$n$.  The following proposition
extends this result to transfinite words.
\begin{prop}\label{pro:ualphavprime}
  Let $u$ and $v$ be two prime words such that $u \ltlex v$.  Then
  $u^\alpha v$ is a prime word for any ordinal~$\alpha$.
\end{prop}
\begin{proof}
  We first prove that every proper suffix~$z$ of~$u^\alpha v$ satisfies
  $u^\alpha v \lelex z$.  Such a suffix~$z$ is either a suffix of~$v$, or
  it has the form $yu^\beta v$ where $y$ is a non-empty suffix of~$u$
  and~$0 \le \beta \le \alpha$.  In the former case, one has
  $u^\alpha v \lelex v$ by Corollary~\ref{cor:ualphav} and $v \lelex z$,
  since $v$ is prime and $z$ is a suffix of~$v$.  In the latter case,
  either $y = u$ or $u \ltstr y$ because $u$ is prime and $y$ is a suffix
  of~$u$.  If $u = y$, then $z = u^{1+\beta}v$ and the result follows from
  Corollary~\ref{cor:ualphav}.  If $u \ltstr y$, then
  $u^\alpha v \ltstr y \lelex yu^\beta v$.

  We now prove that $u^\alpha v$ is primitive.  Suppose that
  $u^\alpha v = z^\beta$ for some primitive word~$z$ and some ordinal
  $\beta \ge 2$.  By Lemma~\ref{lem:power} and by the first paragraph, the
  ordinal $\beta$ is a power of~$\omega$.  Note that
  $u^\alpha = z^{\beta_1}$ and $v = z^{\beta_2}$ is impossible: since $v$
  is primitive, $\beta_2 = 1$, $z = v$ and $u^\alpha = z^{\beta_1}$, which
  contradicts the fact that $u^\alpha \ltlex u^\alpha v \lelex v$.  Then
  there exist two ordinals $\beta_1$ and~$\beta_2$ and two words $z_1$
  and~$z_2$ such that $z = z_1z_2$, $u^\alpha = z^{\beta_1} z_1$ and
  $v = z_2z^{\beta_2}$.

  Since $\beta = \beta_1 + 1 + \beta_2$ and $\beta$ is a power of~$\omega$,
  then $\beta_2 = \beta$.  Since $\beta_2 \ge \omega$, $\beta_2$ can be
  written as $\beta_2 = \omega + \beta'_2$ where $\beta'_2$ is either $0$
  or a limit ordinal.  The word~$v$ is then equal to
  $z_2z^{\omega + \beta'_2}$.  Since $u^\alpha = {(z_1z_2)}^{\beta_1}z_1$,
  the word $z_1$ is a prefix of~$u^\alpha$.  Therefore it satisfies
  $z_1 \lelex u^\alpha$, and since $u^\alpha \ltlex v$ by
  Corollary~\ref{cor:ualphav}, it also satisfies $z_1 \ltlex v$.  If it
  satisfies $z_1 \ltstr v$, the suffix
  $z' = z^{\omega + \beta'_2} = {(z_1z_2)}^\omega z^{\beta'_2}$ of~$v$
  satisfies $z' \ltlex v$, and it contradicts the fact that $v$ is prime.
  It follows that $z_1$ is a prefix of~$v$ and thus a prefix of~$z_2z_1$.
  The equality $z_1 = z_2z_1$ is not possible.  Otherwise, $v$ is equal to
  $z_1^{\omega + \beta'_2}$ and it is not primitive.  Therefore $z_2z_1$ is
  equal to $z_1z_3$ for some non-empty word~$z_3$.  The suffix
  $z' = z_3z_2z^{\omega + \beta'_2}$ of $v$ satisfies $v \lelex z'$.  It
  follows that $z_1v \lelex z_1z' = v$.  Since $z_1v$ is equal to the
  suffix $z^{\omega + \beta'_2}$, the equality $v = z^{\omega + \beta'_2}$
  must hold and $v$ is not primitive.
\end{proof}

\begin{exa}
  Consider again the sequence ${(u_n)}_{n < \omega}$ of words defined by
  $u_0 = a$ and $u_{n+1} = u_n^\omega b$.  It follows from the previous
  result that each word~$u_n$ is prime.
\end{exa}

The following proposition is an extension to transfinite words of the
statement of Proposition~2.2 in~\cite{Siromoney94} that the limit of finite
prime words is a prime $\omega$-word.
\begin{prop}\label{pro:limit}
  Let ${(u_n)}_{n<\omega}$ be an $\omega$-sequence of words such that the
  product $u_0 \cdots u_n$ is prime for each $n < \omega$.  Then the
  $\omega$-product $u_0u_1u_2 \cdots $ is also prime.
\end{prop}
\begin{proof}
  Let $u$ be the $\omega$-product $u_0u_1u_2 \cdots$.  We first prove that
  each suffix~$z$ of~$u$ satisfies $u \lelex z$.  A proper suffix~$z$
  of~$u$ has the form $z = u'_{k}u_{k+1}u_{k+2}\cdots$ where $k < \omega$ and
  $u'_k$ is a non-empty suffix of~$u_k$.  Since $u'_k$ is a non-empty
  suffix of the prime word $u_0 \cdots u_k$, it satisfies either
  $u_0 \cdots u_k = u'_k$ or $u_0 \cdots u_k \ltstr u'_k$.  In both cases,
  the suffix~$z$ satisfies $u \lelex z$.

  We now prove that $u$ is primitive.  Suppose that $u = y^\alpha$ for some
  ordinal~$\alpha \ge 2$.  Since $\alpha \ge 2$, $y$ is a proper prefix
  of~$u$.  There exists then an integer~$k$ such that $y$ is a proper
  prefix of $u_0 \cdots u_k$: $y \ltpre u_0 \cdots u_k$.  Since
  $u = y^\alpha$, there exist an ordinal~$\gamma$ such that
  $u_0 \cdots u_k$ is a prefix of~$y^\gamma$.  Let $\gamma$ be the least
  ordinal such that $u_0 \cdots u_k$ is a prefix of~$y^\gamma$:
  $y \ltpre u_0 \cdots u_k \lepre y^\gamma$.  Since $u_0 \cdots u_k$ is
  primitive, $u_0 \cdots u_k$ is not equal to~$y^\gamma$, that is,
  $y \ltpre u_0 \cdots u_k \ltpre y^\gamma$.  We claim that the
  ordinal~$\gamma$ is a successor ordinal.  For any ordinal
  $\gamma' < \gamma$, $y^{\gamma'}$ is a prefix of~$u$, but
  $u_0 \cdots u_k$ is not a prefix of~$y^{\gamma'}$.  It follows that
  $y^{\gamma'}$ is a prefix of $u_0 \cdots u_k$.  The ordinal~$\gamma$ is
  then a successor ordinal, since the set
  $\Omega = \{ \beta \mid y^\beta \lepre u_0 \cdots u_k \}$ is closed.  The
  word $u_0 \cdots u_k$ is then equal to~$y^{\gamma'}y'$ where
  $\gamma = \gamma' + 1$ and $y'$ is a proper prefix of~$y$.  This
  word~$y'$ is a suffix and a proper prefix of~$u_0 \cdots u_k$ and this
  contradicts the fact that $u_0 \cdots u_k$ is prime.
\end{proof}

\begin{exa}\label{exa:involvedcont}
  Consider once again the sequence ${(u_n)}_{n < \omega}$ of words defined by
  $u_0 = a$ and $u_{n+1} = u_n^\omega b$, and let $u_\omega$ be the word
  $u_0u_1u_2 \cdots$.  Since each word~$u_n$ is prime and since
  $u_0 \cdots u_n = u_n$, the limit word~$u_\omega$ is also prime.
\end{exa}

\begin{lem}\label{lem:ualphavbetaprime}
  Let $u$ and $v$ be two prime words such that $u \ltlex v$ and let
  $\alpha$ be an ordinal such that $u^\alpha v \ltlex v$.  The word
  $u^\alpha v^\beta$ is prime for any ordinal~$\beta \ge 1$.
\end{lem}

Note that the relation $u \ltlex v$ implies $u^\alpha v \lelex v$ by
Corollary~\ref{cor:ualphav}.  Lemma~\ref{lem:ualphavbetaprime} assumes that
$u^\alpha v \neq v$ because otherwise $u^\alpha v^\beta$ is equal to
$v^\beta$ and it is not primitive whenever $\beta \ge 2$.

\begin{proof}
  By Lemma~\ref{lem:vequalphaupw}, the word~$v$ is equal to $u^\gamma xy$
  for some ordinal~$\gamma$ where $|x| \le |u|$ and $u \ltstr x$.  If
  $\alpha + \gamma = \gamma$, then
  $u^\alpha v = u^{\alpha+\gamma}xy = u^\gamma xy$ is equal to~$v$ and this
  is a contradiction with the hypothesis $u^\alpha v \ltlex v$.  Therefore,
  we may assume that $\alpha + \gamma > \gamma$.  Then
  $u^\alpha v^\beta = u^\alpha vv^{\beta'} = u^{\alpha+\gamma}
  xyv^{\beta'}$ where $\beta'$ is either $\beta-1$ is $\beta < \omega$ or
  $\beta$ otherwise.  It follows that $u^\alpha v^\beta \ltlex v$ holds for
  any ordinal $\beta \ge 1$.  The proof that the word~$u^\alpha v^\beta$ is
  prime is then carried out by induction on~$\beta$.  The case $\beta = 1$
  is the result of Proposition~\ref{pro:ualphavprime}.  If $\beta > 1$, the
  result follows again from Proposition~\ref{pro:ualphavprime} if $\beta$
  is a successor ordinal and it follows from Proposition~\ref{pro:limit} if
  $\beta$ is a limit ordinal.
\end{proof}

It can easily be shown by induction on~$n$ that if the finite sequence
$u_1,\ldots,u_n$ of prime words satisfies $u_1 \ltlex \cdots \ltlex u_n$,
then the product $u_1 \cdots u_n$ is still prime.  By
Proposition~\ref{pro:limit}, this is also true for a sequence of
length~$\omega$.  This no longer holds for longer sequences.  Consider
again the sequence ${(u_\alpha)}_{\alpha \le \omega}$ of length $\omega+1$ of
prime words given in Example~\ref{exa:involvedcont}.  Their product
$\prod_{\alpha \le \omega}{u_\alpha}$ is equal to~$u_\omega^2$ and it is
not prime.

\section{Factorization in prime words}\label{sec:factorization}

In this section, we prove that any word has a unique factorization into
prime words that is almost non-increasing.  The goal is to extend to
transfinite words the classical result that any finite word is the product
of a non-increasing sequence of prime words~\cite[Thm~5.1.5]{Lothaire}.  It
turns out that this extension is not straightforward, since some words are
not equal to a product of a non-increasing sequence of prime words.  Let us
consider the $\omega$-word $x = aba^2ba^3 \cdots$ and the $(\omega+1)$-word
$xb$.  The word~$x$ can be factorized as
$x = ab \cdot a^2b \cdot a^3b \cdots$ and the sequence ${(a^{n}b)}_{n<\omega}$
is indeed a non-increasing sequence of prime words.  The word $xb$,
however, cannot be factorized into a non-increasing sequence of prime
words.  A naive attempt could be $ab \cdot a^2b \cdot a^3b \cdots b$, but
the sequence ${(u_n)}_{n\le\omega}$ where $u_n = a^{n+1}b$ for $n<\omega$ and
$u_\omega = b$ is not non-increasing since $u_n \ltlex u_\omega$ for each
$n < \omega$.  This naive attempt is the only possible one since, for
finite words as well for $\omega$-words~\cite{Siromoney94}, the first
factor is always the longest prime prefix.  This property also holds in our
case (see Proposition~\ref{pro:longestprefix}).  To cope with this
difficulty, we introduce the notion of a densely non-increasing sequence.
This is a slightly weaker notion than the notion of a non-increasing
sequence.  A densely non-increasing sequence ${(u_\beta)}_{\beta<\alpha}$ may
have some $\gamma < \gamma' < \alpha$ such that
$u_{\gamma} \ltlex u_{\gamma'}$, but this may only happen if there exists a
limit ordinal $\gamma < \gamma'' \le \gamma'$ such that the sequence
${(u_\beta)}_{\beta<\alpha}$ is cofinally decreasing in~$\gamma''$.  Roughly
speaking, an increase is allowed if it comes after an $\omega$-sequence of
strict decreases.  The $(\omega+1)$-sequence ${(u_n)}_{n\le\omega}$ where
$u_n = a^{n}b$ for $n<\omega$ and $u_\omega = b$ is densely non-increasing.
Indeed, one has $u_n \ltlex u_\omega$, but also $u_n \gtlex u_{n+1}$ for
each $n < \omega$.

We now introduce the formal definition of a densely non-increasing
sequence.  We only use this notion for sequences of prime words
lexicographically ordered, but we give the definition for an arbitrary
ordered set~$U$.  Let $(U,<)$ be a linear ordering and let
$\bar{u} = {(u_\beta)}_{\beta<\alpha}$ be a sequence of elements of~$U$.  The
sequence $\bar{u}$ is \emph{constant} in the interval $\ropen{\gamma,\gamma'}$
where $\gamma < \gamma' \le \alpha$ if $u_\beta = u_\gamma$ holds for any
$\gamma \le \beta < \gamma'$.  As usual, the sequence~$x$ is
\emph{non-increasing} if for any $\beta$ and~$\beta'$,
$\beta < \beta'< \alpha$ implies $u_\beta \ge u_{\beta'}$.

\begin{defi}
  It is \emph{densely non-increasing} if for any interval
  $\ropen{\gamma, \gamma'}$ where $\gamma < \gamma' \le \alpha$, either it is
  constant in $\ropen{\gamma,\gamma'}$ or there exist two ordinals
  $\gamma \le \beta < \beta' < \gamma'$ such that $u_\beta > u_{\beta'}$.
\end{defi}
It is clear that a non-increasing sequence is also densely non-increasing.
The converse does not hold as it is shown by the already considered
$(\omega+1)$-sequence ${(u_\beta)}_{\beta \le \omega}$ defined by
$u_n = a^{n}b$ for $n < \omega$ and $u_\omega = b$.  The following
proposition provides a characterization of densely non-increasing
sequences.  It also gives some insight on the property of being densely
non-increasing.

\begin{prop}\label{pro:charlocnonincr}

  The sequence $\bar{u} = {(u_\beta)}_{\beta<\alpha}$ is densely
  non-increasing if and only if the following two statements hold for any
  ordinals $\beta' < \beta < \alpha$.
  \begin{itemize} \itemsep0cm
  \item If $\beta = \beta' + 1$, then $u_{\beta'} \ge u_\beta$.
  \item If $\beta$ is a limit ordinal and $\bar{u}$ is constant in
    $\ropen{\beta',\beta}$, then $u_{\beta'} \ge u_\beta$.
  \end{itemize}
\end{prop}
\begin{proof}
  Applying the definition of densely non-increasing to the interval
  $[\beta', \beta]$ gives that the two statements are obviously necessary.

  Conversely we prove by transfinite induction on~$\beta$ that if the two
  hypothesis are satisfied, then the restriction of~$\bar{u}$ to the
  interval $[0,\beta]$ is densely non-increasing.  The case $\beta = 0$ is
  trivially true.  The case $\beta = \beta'+1$ is handled by the first hypothesis
  and the case $\beta$ being a limit ordinal is handled by the second
  hypothesis.
\end{proof}

As pointed out by a referee, a sequence is densely non-increasing if and
only if it is non-increasing on any interval where it is monotone.

The following theorem is the main result of the paper.  It extends the
classical result that states that any finite word can be uniquely written
as a non-increasing product of prime words~\cite[Thm~5.1.5]{Lothaire}.  A
\emph{prime factorization} of a word~$x$ is a densely non-increasing
sequence ${(u_\beta)}_{\beta<\alpha}$ of prime words such that
$x = \prod_{\beta<\alpha}{u_\beta}$.
\begin{thm}\label{thm:main}
  For any word $x \in A^{\#}$, there exists a unique \emph{prime
    factorization} of~$x$.
\end{thm}

\begin{exa}
  The prime factorization of the finite words $aabab$ and $abaab$ are
  $aabab$ and $ab \cdot aab$ since $ab$, $aab$ and $aabab$ are prime words.
  The prime factorization of the $\omega$-words $x_0 = aba^2ba^3b \cdots$
  and $x_1 = abab^2ab^3 \cdots$ are $x_0 = ab \cdot a^2b \cdot a^3b \cdots$
  and $x_1 = abab^2ab^3 \cdots$ since $ab,a^2b, a^3b,\ldots$ and
  $x_1 = abab^2ab^3 \cdots$ are prime words.

  The prime factorization of the $(\omega+1)$-word $x_2 = x_0b$ is the
  $(\omega+1)$-sequence ${(u_\beta)}_{\beta \le \omega}$ given by
  $u_n = a^{n+1}b$ for $n < \omega$ and $u_\omega = b$.  This factorization
  is not non-increasing since $u_0 = ab \ltlex b = u_\omega$, but it is
  densely non-increasing.
\end{exa}

The proof of the theorem is organized as follows.  In the next section, we
give a few properties of densely non-increasing sequences.  These
properties are used in the next two sections.  We prove in
Section~\ref{sec:existence} that the factorization in prime words
always exists and we prove in Section~\ref{sec:uniqueness} that it is
unique.  Surprisingly, the uniqueness is useful in one of the proofs of the
existence.

\subsection{Properties of densely non-increasing sequences}

In this section, we establish a few properties of densely non-increasing
sequences that are needed for the proof of Theorem~\ref{thm:main}.  In this
section, all sequences are formed of elements from an arbitrary ordered
set~$U$.

\begin{defi}
  Let $\bar{u} = {(u_\beta)}_{\beta<\alpha}$ be a sequence and let $\gamma$
  be a limit ordinal such that $\gamma \le \alpha$.  The sequence~$\bar{u}$
  is \emph{ultimately constant} in~$\gamma$ if there exists
  $\gamma' < \gamma$ such that it is constant in the interval
  $\ropen{\gamma', \gamma}$.
\end{defi}
If the sequence~$\bar{u}$ is densely non-increasing but not ultimately
constant in~$\gamma$, then for any $\gamma' < \gamma$, there exist two
ordinals $\beta$ and~$\beta'$ such that
$\gamma' \le \beta < \beta' < \gamma$ and $u_\beta > u_{\beta'}$.

Any sequence has a longest prefix that is non-decreasing.  The following
lemma states that when the sequence is densely non-increasing, but not
non-increasing, the length of this longest prefix is a limit ordinal and
the sequence is not ultimately constant at this ordinal.
\begin{lem}\label{lem:decrprefix}
  Let $\bar{u} = {(u_\beta)}_{\beta<\alpha}$ be a densely non-increasing
  sequence.  If $\bar{u}$ is not non-increasing, there exists a greatest
  ordinal $\alpha' < \alpha$ such that ${(u_\beta)}_{\beta<\alpha'}$ is
  non-increasing.  Furthermore, this ordinal~$\alpha'$ is a limit ordinal
  and the sequence~$\bar{u}$ is not ultimately constant in~$\alpha'$.
\end{lem}
\begin{proof}
  Let $\Omega$ be the set
  $\{ \gamma \le \alpha \mid {(u_\beta)}_{\beta<\gamma} \text{ is
    non-increasing}\}$.  Since this set of ordinals is closed, it has a
  greatest element~$\alpha'$ that is strictly smaller than~$\alpha$ since
  $\bar{u}$ is not non-increasing.  We claim that this ordinal~$\alpha'$ is
  a limit ordinal.  Suppose, by contradiction, that $\alpha'$ is a
  successor ordinal: $\alpha' = \alpha'' + 1$.  Since $\bar{u}$ is densely
  non-increasing, one has $u_{\alpha''} \ge u_{\alpha'}$ and this is a
  contradiction since $\alpha'+1$ should belong to~$\Omega$.  We now prove
  that the sequence~$\bar{u}$ is not ultimately constant in~$\alpha'$.
  Suppose again, by contradiction, that the sequence $\bar{u}$ is
  ultimately constant in~$\alpha'$.  There exists an ordinal
  $\gamma<\alpha'$ such that $u_\beta = u_\gamma$ for any
  $\gamma \le \beta < \alpha'$.  If $u_\gamma < u_{\alpha'}$, the sequence
  $\bar{u}$ is not densely non-increasing.  Therefore
  $u_{\alpha'} \le u_\gamma$ and this is again a contradiction since
  $\alpha'+1$ should again belong to~$\Omega$.
\end{proof}

The \emph{range} of a sequence $\bar{u} = {(u_\beta)}_{\beta<\alpha}$ is the
set of values that occur in the sequence.  More formally, it is the set
$\{ u_\beta \mid \beta<\alpha \}$.

\begin{cor}\label{cor:finiterange}
  Let $\bar{u} = {(u_\beta)}_{\beta<\alpha}$ be a densely non-increasing
  sequence.  If the range of~$\bar{u}$ is finite, it is non-increasing.
\end{cor}
\begin{proof}
  Suppose that the sequence~$\bar{u}$ is not constant.  By
  Lemma~\ref{lem:decrprefix}, there exists a greatest
  ordinal~$\alpha' < \alpha$ such that
  $\bar{u}' = {(u_\beta)}_{\beta<\alpha'}$ is non-increasing.  Furthermore,
  the sequence~$\bar{u}'$ is not ultimately constant in~$\alpha'$.  This
  implies that the range of $\bar{u}'$ is infinite.
\end{proof}

\begin{lem}
  Let $\bar{u} = {(u_\beta)}_{\beta<\alpha}$ be a sequence.  If the range
  of~$\bar{u}$ is infinite, there exists a limit ordinal
  $\alpha' \le \alpha$ such that the sequence~$\bar{u}$ is not ultimately
  constant in~$\alpha'$.
\end{lem}
\begin{proof}
  Let $\alpha'$ be the least ordinal such that the range of
  ${(u_\beta)}_{\beta<\alpha'}$ is infinite.  This ordinal~$\alpha'$ is a
  limit ordinal: indeed, if it is a successor ordinal, that is,
  $\alpha' = \alpha'' + 1$, the range of the sequence
  ${(u_\beta)}_{\beta<\alpha''}$ is still infinite and this is a
  contradiction with the definition of~$\alpha'$.  It is also clear that
  $\bar{u}$ cannot be ultimately constant in~$\alpha'$.  Indeed, if
  $\bar{u}$ is ultimately constant in~$\alpha'$, there exists an ordinal
  $\alpha'' < \alpha$ such that the range of the sequence
  ${(u_\beta)}_{\beta<\alpha''}$ is still infinite and this is again a
  contradiction with the definition of~$\alpha'$.
\end{proof}

\begin{lem}\label{lem:factfinite}
  Let $\bar{u} = {(u_\beta)}_{\beta<\alpha}$ be a densely non-increasing
  sequence.  If the range of~$\bar{u}$ is infinite, it can be uniquely
  factorized as $\bar{u} = \bar{v}\bar{w}$ where the length~$\gamma$
  of~$\bar{v}$ is a limit ordinal, $\bar{v}$ is not ultimately constant
  in~$\gamma$ and the range of~$\bar{w}$ is finite.
\end{lem}
Note that if the range of~$\bar{u}$ is finite, it also has a degenerate
factorization $\bar{u} = \bar{v}\bar{w}$ where $\bar{v}$ is the empty
sequence and $\bar{w} = \bar{u}$ has a finite range.

\begin{proof}
  Let $\Omega$ be the set of limit ordinals given by
  \begin{displaymath}
    \Omega = \{ \alpha' \le \alpha \mid \text{$\alpha'$ limit ordinal and $\bar{u}$ is not ultimately constant in~$\alpha'$}\}.
  \end{displaymath}
  By the previous lemma, the set~$\Omega$ is non-empty.  We claim that it
  is closed. Suppose $\beta = \sup \{ \beta_n | n < \omega\}$ where each
  $\beta_n$ is an element of~$\Omega$.  Since each $\beta_n$ is a limit
  ordinal, so is $\beta$.  Suppose by contradiction that $\beta$ does not
  belong to~$\Omega$.  The sequence $\bar{u}$ is thus ultimately constant
  in~$\beta$.  There is an interval to the left of~$\beta$ where $\bar{u}$
  is constant.  Each ordinal in this interval is not in~$\Omega$ and this
  is a contradiction with the definition of~$\beta$.  Hence $\Omega$ is
  closed and let $\gamma$ be its greatest element.  Let $\bar{v}$ be the
  sequence ${(u_\beta)}_{\beta<\gamma}$ and let $\bar{w}$ be the unique
  sequence such that $\bar{u} = \bar{v}\bar{w}$.  The length of~$\bar{v}$
  is the limit ordinal~$\gamma$ and the sequence~$\bar{v}$ is not
  ultimately constant in~$\alpha'$.  It remains to prove that the range
  of~$\bar{w}$ is finite.  If the range of~$\bar{w}$ is infinite, there
  exists, by the previous lemma, a limit ordinal~$\gamma'$ where $\bar{w}$
  is not ultimately constant.  This contradicts the definition of~$\gamma$.
  The factorization is unique since $\gamma$ must be the greatest limit
  ordinal where $\bar{u}$ is not ultimately constant.
\end{proof}

\subsection{Existence of the factorization}\label{sec:existence}

We prove in this section that any transfinite word has a prime
factorization.  We actually give two proofs of the existence of the prime
factorization.  The first one is based on Zorn's lemma and the second one
uses a transfinite induction on the length of words.  The former one is
shorter, but the latter one provides a much better insight.  The latter one
needs the uniqueness of the factorization.  The proof of this uniqueness is
given in the next section and it does not use the existence.  We
first sketch the proof based on Zorn's lemma and then we detail the proof
by transfinite induction.

We now sketch the proof based on Zorn's lemma. Let $x$ be a fixed word.
Let $X$ be the set of sequences $\bar{u} = {(u_\beta)}_{\beta<\alpha}$ of
prime words such that $x = \prod_{\beta<\alpha}{u_\beta}$.  Note that it is
not assumed that the sequence~$\bar{u}$ is densely non-increasing.  We
define an ordering~$<$ on the sequences of words as follows.  Two sequences
$\bar{u} = {(u_\beta)}_{\beta<\alpha}$ and
$\bar{u}' = {(u'_\beta)}_{\beta<\alpha'}$ satisfy $\bar{u} < \bar{u}'$ if
$\bar{u}$ refines $\bar{u}'$.  This means that there exists a sequence
${(\gamma_\beta)}_{\beta<\alpha'}$ of ordinals such that
$u'_\beta = \prod_{\gamma_\beta \le \eta < \gamma_{\beta+1}}{u_\eta}$ for
each $\beta < \alpha'$.  For any totally ordered non-empty subset~$Y$
of~$X$, there exists a least upper bound
$\bar{u} = {(u_\beta)}_{\beta<\alpha}$ which is constructed as follows.  Two
symbols $a_\gamma$ and $a_{\gamma'}$ of~$x$ end up in the same factor
$u_\beta$ as soon as they are in the same factor of at least one
factorization in~$Y$.  It can be observed that each word~$u_\beta$ either
occurs in some sequence of~$Y$ or is the limit of words occurring in
sequences of~$Y$.  In the former case, the word~$u_\beta$ is prime by
definition of~$X$ and in the latter case, it is prime by
Proposition~\ref{pro:limit}.  This shows that each word~$u_\beta$ is prime
and that the sequence $\bar{u} = {(u_\beta)}_{\beta<\alpha}$ belongs to~$X$.
This allows us to apply Zorn's lemma: the set~$X$ has a maximal element
$\bar{v} = {(v_\beta)}_{\beta<\alpha}$.  It remains to show that this
sequence~$\bar{v}$ is indeed densely non-increasing.  Suppose by
contradiction that it is not. By Proposition~\ref{pro:charlocnonincr},
there is an ordinal~$\beta$ such that either $\beta = \beta'+1$ and
$v_{\beta'} \ltlex v_\beta$ or $\beta$ is a limit ordinal where $\bar{v}$
is constant in $\ropen{\beta',\beta}$ and $v_{\beta'} \ltlex v_\beta$.  In the
former case, $v_{\beta'}v_\beta$ is prime by
Proposition~\ref{pro:ualphavprime} and in the latter case
$v_{\beta'}^{\beta-\beta'}v_\beta$ is also prime by
Proposition~\ref{pro:ualphavprime}.  In both cases, this is a contradiction
with the maximality of~$\bar{v}$.

We now give the second proof of the existence.  We start with an easy lemma
on ordinals which states that any sequence of ordinals contains a
non-decreasing sub-sequence.  It is used in the proof of the main result of
this section, namely Proposition~\ref{pro:existsfact}.

\begin{lem}\label{lem:incr-seq-ordinals}
  For any sequence ${(\alpha_n)}_{n < \omega}$ of ordinals, there exists a
  non-decreasing sub-sequence ${(\alpha_{k_n})}_{n < \omega}$ (where
  ${(k_n)}_{n<\omega}$ is an increasing sequence of integers).
\end{lem}
The proof follows from the fact that the ordering of countable ordinals is
a linear well quasi ordering.

The following lemma is an easy consequence of Corollary~\ref{cor:ualphav} and
Lemma~\ref{lem:ualphavbetaprime}.  It is stated because the same reasoning
is used several time.
\begin{lem}\label{lem:concatuv}
  Let $u$ and~$v$ be two prime words such that $u \lelex v$ and let
  $\alpha$ and~$\beta$ be two non-zero ordinals.  The word
  $u^\alpha v^\beta$ is equal to $w^\gamma$ where the word~$w$ is prime and
  $\gamma$ is an ordinal.  Furthermore, either $w = v$ and
  $\gamma \in \{\beta, \alpha+\beta\}$ or $w = u^\alpha v^\beta$ and
  $\gamma = 1$.
\end{lem}
\begin{proof}
  If $u = v$, the word $u^\alpha v^\beta$ is equal to $v^{\alpha+\beta}$:
  set $w = v$ and $\gamma = \alpha + \beta$.  We now suppose that
  $u \ltlex v$ and thus $u^\alpha v \lelex v$ by
  Corollary~\ref{cor:ualphav}.  If $u^\alpha v = v$, the word
  $u^\alpha v^\beta$ is equal to $v^\beta$: set $w = v$ and
  $\gamma = \beta$.  We finally suppose that $u^\alpha v \ltlex v$.  The
  word $u^\alpha v^\beta$ is prime by Lemma~\ref{lem:ualphavbetaprime}: set
  $w = u^\alpha v^\beta$ and $\gamma = 1$.
\end{proof}

The following lemma is obtained by repeatedly applying
Lemma~\ref{lem:concatuv}.  This lemma states that if a word is already
factorized as powers of prime words, its prime factorization is obtained
by grouping these powers of prime words using Lemma~\ref{lem:concatuv}.
\begin{lem}\label{lem:finitefact}
  A word $x = u_1^{\alpha_1} \cdots u_m^{\alpha_m}$ where each word~$u_i$
  is prime and each $\alpha_i$ is an ordinal, has a prime factorization
  $x = v_1^{\beta_1} \cdots v_n^{\beta_n}$ where $m \ge n$,
  $v_1 \gtlex \cdots \gtlex v_n$ and each prime word~$v_j$ is either a
  word~$u_i$ or a product $u_i^{\alpha_i} \cdots u_k^{\alpha_k}$ for
  $1 \le i < k \le n$.
\end{lem}
 \begin{proof}
   The proof is by induction on the integer~$m$.  The result is clear if
   $m = 1$: just set $n = 1$ and $v_1 = u_1$.  The result is also clear if
   $u_1 \gtlex \cdots \gtlex u_m$: just set $n = m$ and $v_i = u_i$ for
   $1 \le i \le n$.  We now suppose that there exists an integer
   $1 \le i < m$ such that $u_i \lelex u_{i+1}$.  By
   Lemma~\ref{lem:concatuv}, the word
   $u_i^{\alpha_i} u_{i+1}^{\alpha_{i+1}}$ is equal to $w^\gamma$ for some
   prime word~$w$ and some ordinal~$\gamma$.  The word~$x$ is then equal to
   $u_1^{\alpha_1} \cdots u_{i-1}^{\alpha_{i-1}} w^\gamma
   u_{i+2}^{\alpha_{i+2}} \cdots u_m^{\alpha_m}$ and the result follows
   from the induction hypothesis.
\end{proof}

A slightly different version of Lemma~\ref{lem:finitefact} is stated below
although it is not needed until Appendix~\ref{sec:rational}.  Its proof is a
straightforward adaptation of the proof given just above.
\begin{lem}\label{lem:factproduct}
  Given prime factorizations of two words
  $x = u_1^{\alpha_1} \cdots u_m^{\alpha_m}$ and
  $y = v_1^{\beta_1} \cdots v_n^{\beta_n}$.  Then $xy$ has a prime
  factorization of the form
  $xy = u_1^{\alpha_1} \cdots u_i^{\alpha_i}w^\gamma v_j^{\beta_j} \cdots
  v_n^{\beta_n}$ where $1 \le i \le m$, $1 \le j \le n$ and $w$ is a prime
  word.
\end{lem}

The following proposition states that any word has a prime factorization.
\begin{prop}\label{pro:existsfact}
  For any word $x \in A^{\#}$, there exists a densely non-increasing
  sequence ${(u_\beta)}_{\beta<\alpha}$ of prime words such that
  $x = \prod_{\beta<\alpha}{u_\beta}$.
\end{prop}

Before giving the formal proof of the proposition, we give a sketch of the
proof.  The existence of the factorization of $x$ is proved by induction on
the length of~$x$. If this length is a successor ordinal (case A in the
formal proof), the result follows by directly using some lemmas given
previously in the paper. The difficult part turns out to be when the length
of~$x$ is a limit ordinal (case B in the formal proof below). This case
requires the uniqueness of the factorization proved in the next section. In
order to help the reader, we roughly describe what is going on in this
case.

First we work with a sequence of words $x_n$ converging to $x$; these words
have increasing lengths $\gamma_n$ converging to the length~$\gamma$
of~$x$. The induction hypothesis allows the use of the factorizations of
each $x_n$ and the aim is to show how these factorizations, in some sense,
converge to the desired factorization of~$x$. To make clear the difficulty,
consider first $x = a^\omega$. Suppose then that $x_n =a^n$. The
factorization of $x_n$ is exactly $a.a. \dots a$ and it gives rise to the
desired factorization of $x$ in $a.a. \dots a. \dots$.

Now consider $x = ab^\omega$ and suppose $x_n = ab^n$. The factorization of
$x_n$ reduces to a single factor $x_n$ itself. Then the limit obtained is
$x$.

This shows that two different situations may occur:

We first compute the number of factors in the factorization of~$x_n$ and
get the supremum of these numbers~$\alpha$.

Situation 1: when we fix any ordinal number of factors in the
factorizations of the words $x_n$ (less than $\alpha$ and for a large
enough ordinal $n$), the length of the prefix obtained by this number of
factors is always the same (case B1 of the formal proof). This is what
happens for $x = a^\omega$: if we fix the number of factors to $k$, we
cover a prefix of length $k$ for all words $x_n$. Then we show that these
first factors are always the same and that the factorization of~$x$ is
obtained by concatenating all these factors.

Situation 2: we can find an ordinal $\beta \le \alpha$, such that when we
fix the number of factors to $\beta$ in the factorization of the words
$x_n$ (for a large enough ordinal $n$), the lengths of the prefixes
obtained increase as $n$ increases (case B2 of the formal proof). First we
prove that this length grows and converges to~$\gamma$. Then this case
splits again into two sub-cases:

\begin{itemize} \itemsep0cm
\item $\beta$ is a limit ordinal (case B2a of the formal proof)
\item $\beta$ is a successor ordinal (case B2b of the formal proof); this
  is what happens for $ab^\omega$ where $\beta = 1 \le \alpha = 1$ and the
  length of the covered prefix of $x_n$ is $(n+1)$.
\end{itemize}
In each sub-case, we describe the limit factorization obtained for~$x$.

We now turn to the formal proof of Proposition~\ref{pro:existsfact}.
\begin{proof}
  The proof is by induction on the length~$|x|$ of~$x$.  The result is
  obvious if $|x| = 1$, that is, if $x = a$ for some letter~$a$ since $a$
  is a prime word.  We now suppose that the length $|x|$ of~$x$ satisfies
  $|x| \ge 2$.  We distinguish two cases depending on whether $|x|$ is a
  successor or a limit ordinal.

  \paragraph{Case A:\@ $|x|$ is a successor ordinal.}
  We first suppose that $|x|$ is a successor ordinal~$\gamma+1$.  The
  word~$x$ is then equal to~$x'a$ where $x'$ is a word of length~$\gamma$
  and $a$ is a letter.  By the induction hypothesis, there exists a densely
  non-increasing sequence ${(u_\beta)}_{\beta<\alpha'}$ of prime words such
  that $x' = \prod_{\beta<\alpha'}{u_\beta}$.  We distinguish then two
  sub-cases depending on whether the range of this sequence is finite or
  infinite.

  If the range of ${(u_\beta)}_{\beta<\alpha'}$ is finite, this sequence is
  non-increasing by Corollary~\ref{cor:finiterange} and the result follows
  then from Lemma~\ref{lem:finitefact}.

  If the range of the sequence $y = {(u_\beta)}_{\beta<\alpha'}$ is infinite,
  it can be decomposed, by Lemma~\ref{lem:factfinite}, as the concatenation
  of two sequences $y_1$ and~$y_2$ where $y_1$ has length~$\delta$ which is
  a limit ordinal and where it is not ultimately constant and the range
  of~$y_2$ is finite.  This decomposition $y = y_1y_2$ corresponds to a
  factorization $x' = x_1x_2$.  By Lemma~\ref{lem:finitefact}, there exists
  a non-increasing sequence of prime words~$y'_2$ whose product is the
  word~$x_2a$.  Since the sequence~$y_1$ is not ultimately constant in
  $\delta = |y_1|$, the sequence~$y_1y'_2$ is also densely non-increasing.
  This sequence is a prime factorization of the word $x = x'a$.

  \paragraph{Case B:\@ $|x|$ is a limit ordinal.}
  We now suppose that $|x|$ is a limit ordinal~$\gamma$.  There exists then
  an increasing sequence ${(\gamma_n)}_{n<\omega}$ of ordinals such that
  $\gamma = \sup_n \gamma_n$.  Let $x_n$ be the prefix of~$x$ of
  length~$\gamma_n$.  By the induction hypothesis, there exists, for each
  integer~$n$, a densely non-increasing sequence
  ${(u_{n,\beta})}_{\beta<\alpha_n}$ of prime words such that
  $x_n = \prod_{\beta<\alpha_n}{u_{n,\beta}}$.  By
  Lemma~\ref{lem:incr-seq-ordinals}, we may suppose that the sequence
  ${(\alpha_n)}_{n<\omega}$ is non-decreasing.  Let $\alpha$ be the ordinal
  $\sup_n \alpha_n$.  By definition of~$\alpha$, there exists, for any
  ordinal $\beta < \alpha$, an integer~$N$ such that
  $\alpha_{N+1} > \beta$.  Note that $n > N$ implies $\alpha_n > \beta$
  since the sequence ${(\alpha_k)}_{k<\omega}$ is non-decreasing.  We let
  $N_\beta$ denote the least integer such that $\alpha_n > \beta$ holds for
  any $n > N_\beta$.  Note that if $\beta < \beta' < \alpha$, then
  $N_\beta \le N_{\beta'}$.  For $n > N_\beta$, the prime factorization of
  $x_n$ has length $\alpha_n \ge \beta$.  This means that the factor
  $u_{n,\beta}$ exists for $n > N_\beta$. For any $\beta < \alpha$ and any
  $N_\beta < n < \omega$, define the ordinal $\lambda_{n,\beta}$ by
  $\lambda_{n,\beta} = \sum_{\beta'<\beta}{|u_{n,\beta'}|}$.  The ordinal
  $\lambda_{n,\beta}$ is the length of the prefix of~$x$ covered by the
  first $\beta$ factors of the prime factorization of~$x_n$.  Note that
  $\lambda_{n,\beta} \le \gamma_n = |x_n|$ and that the equality
  $\lambda_{n,\beta} = \gamma_n$ holds whenever $\beta = \alpha_n$.  Note
  that the sequence ${(u_{n,\beta'})}_{\beta'<\beta}$ is a prime
  factorization of the prefix $x\ropen{0,\lambda_{n,\beta}}$ of~$x$ which is also
  a prefix of~$x_n$ since $\lambda_{n,\beta} < \gamma_n$.

  We claim that the ordinals $\lambda_{n,\beta}$ have the following two
  properties.  Let $\beta<\alpha$ be an ordinal and let $m$ and~$n$ be two
  integers such that $N_\beta<m<n$.
  \begin{enumerate}[(i)] \itemsep0cm
  \item If $\lambda_{m,\beta} = \lambda_{n,\beta}$, the equality $\lambda_{m,\beta'} = \lambda_{n,\beta'}$ also
    holds for any $\beta' < \beta$.
  \item If $\lambda_{m,\beta} \neq \lambda_{n,\beta}$, then $\lambda_{m,\beta} < \gamma_m < \lambda_{n,\beta}$.
  \end{enumerate}

  \noindent
  We first prove Claim~(i).  Suppose that the equality
  $\lambda_{m,\beta} = \lambda_{n,\beta}$ holds.  The sequences
  ${(u_{m,\delta})}_{\delta<\beta}$ and ${(u_{n,\delta})}_{\delta<\beta}$ are
  two densely non-increasing sequences of prime words.  If
  $\lambda_{m,\beta} = \lambda_{n,\beta}$, their products are equal to the
  prefix $x\ropen{0,\lambda_{m,\beta}}$ of length~$\lambda_{m,\beta}$.  Since
  this factorization is unique by Corollary~\ref{cor:uniquefact} (see
  below), the sequences must coincide and this proves
  $\lambda_{m,\beta'} = \lambda_{n,\beta'}$ for each $\beta' < \beta$.
  This proves Claim~(i).

  Now we prove Claim~(ii).  Note that the relation
  $\lambda_{m,\beta} < \gamma_m$ always holds by definition
  of~$\lambda_{m,\beta}$.  It remains to show that
  $\gamma_m < \lambda_{n,\beta}$.  If
  $\lambda_{m,\beta} \neq \lambda_{n,\beta}$, there exists $\beta' < \beta$
  such that $u_{m,\beta'} \neq u_{n,\beta'}$.  Let $\beta'$ be the least
  ordinal such that $u_{m,\beta'} \neq u_{n,\beta'}$.  By definition
  of~$\beta'$, one has $\lambda_{m,\beta'} = \lambda_{n,\beta'}$.  Since
  $\gamma_m < \gamma_n$, the word~$x_m$ is a prefix of~$x_n$: the
  word~$x_n$ is equal to~$x_{m}z$ for some word~$z$.  By
  Proposition~\ref{pro:longestprefix}, the words $u_{m,\beta'}$
  and~$u_{n,\beta'}$ are respectively the longest prefix of the suffix
  $x_m\ropen{\lambda_{m,\beta'},\gamma_m}$ $x_n\ropen{\lambda_{n,\beta'},\gamma_n}$ of
  $x_m$ and~$x_n$ starting at
  position~$\lambda_{m,\beta'} = \lambda_{n,\beta'}$.  If they are not
  equal, $u_{n,\beta'}$ cannot be a prefix of~$x_m$.  This shows that
  $\lambda_{n,\beta'} + |u_{n,\beta'}| = \lambda_{n,\beta'+1} > |x_m| =
  \gamma_m$.  It follows that
  $\lambda_{n,\beta} \ge \lambda_{n,\beta'+1} > \gamma_m$ since
  $\beta \ge \beta'+1$.  This proves Claim~(ii).

  Let $\beta$ be an ordinal such that $\beta < \alpha$.  The
  ordinals~$\lambda_{n,\beta}$ are defined for any $n > N_\beta$.  Note
  that claim~(ii) implies that the sequence
  ${(\lambda_{n,\beta})}_{n<\omega}$ is non-decreasing.  By Claim~(i), also
  the sequence ${(\lambda_{n,\beta'})}_{n<\omega}$ is ultimately constant
  in~$\omega$ for any ordinal $\beta' < \beta$.  If no such integer
  $N'_\beta$ exists, for any integer~$n$, there exists, by Claim~(ii), an
  integer~$m$ such that $\lambda_{m,\beta} \ge \gamma_n$.  Thus, the
  sequence ${(\lambda_{n,\beta})}_{n<\omega}$ converges to~$\gamma$ when $n$
  goes to~$\omega$.  We distinguish two sub-cases depending on whether
  there exists, or not, an ordinal~$\beta < \alpha$ such that
  ${(\lambda_{n,\beta})}_{n<\omega}$ is not ultimately constant in~$\omega$.

  \paragraph{Case B1:} We first suppose that for each~$\beta<\alpha$, the
  sequence ${(\lambda_{n,\beta})}_{n<\omega}$ is ultimately constant
  in~$\omega$.  For each $\beta<\alpha$, there exists an integer~$N'_\beta$
  and an ordinal~$\lambda_\beta$ such that, for any $n > N'_\beta$,
  $\lambda_{n,\beta} = \lambda_\beta$.  For any $\beta' < \beta < \alpha$,
  it follows from $\lambda_{n,\beta'} < \lambda_{n,\beta}$ for each
  $n > N_\beta$ that $\lambda_{\beta'} < \lambda_{\beta}$.  Let
  ${(u_\beta)}_{\beta<\alpha}$ be the sequence of words defined by
  $u_\beta = x\ropen{\lambda_\beta,\lambda_{\beta+1}}$.  We claim that the
  sequence ${(u_\beta)}_{\beta<\alpha}$ is a prime factorization of~$x$.  We
  first prove that $\sup_{\beta} \lambda_\beta = \gamma = |x|$.  Let
  $\delta$ be an ordinal such that $\delta < \gamma$.  The result is
  obtained as soon as there exists $\beta < \alpha$ such that
  $\lambda_\beta > \delta$.  Since $\gamma = \sup_n \gamma_n$, there exists
  an integer~$n$ such that $|x_n| = \gamma_n > \delta$.  Since
  $|x_n| > \delta$, there exists an ordinal $\beta \le \alpha_n$ such that
  $\lambda_{n,\beta} > \delta$.  Since the sequence
  ${(\lambda_{n,\beta})}_{n<\omega}$ is non-decreasing, one has
  $\lambda_\beta > \delta$.  This proves that
  $\sup_{\beta} \lambda_\beta = \gamma = |x|$ and that
  ${(u_\beta)}_{\beta<\alpha}$ is indeed a factorization of~$x$.  For
  $n > N'_\beta$, one has, by Claim~(i),
  $\lambda_{n,\beta'} = \lambda_{\beta'}$ for any $\beta' < \beta$ and thus
  $u_{n,\beta'} = u_{\beta'}$.  This means that the sequence
  ${(u_{\beta'})}_{\beta'<\beta}$ is a prime factorization of the prefix
  $x\ropen{0,\lambda_\beta}$.  Since this is true for each $\beta < \alpha$, the
  sequence ${(u_{\beta'})}_{\beta'<\alpha}$ is a prime factorization of~$x$.

  \paragraph{Case B2:} We now suppose that there exists, at least, one
  ordinal $\beta<\alpha$ such that the sequence
  ${(\lambda_{n,\beta})}_{n<\omega}$ is not ultimately constant in~$\omega$.
  Let $\beta$ be the least ordinal such that
  ${(\lambda_{n,\beta})}_{n<\omega}$ is not ultimately constant in~$\omega$.
  Note that $\beta > 0$ since $\lambda_{n,0} = 0$ for any $n < \omega$ and
  the sequence ${(\lambda_{n,0})}_{n<\omega}$ is ultimately constant
  in~$\omega$.  By definition of~$\beta$, for each $\beta'<\beta$, the
  sequence ${(\lambda_{n,\beta'})}_{n<\omega}$ is ultimately constant
  in~$\omega$: there exists an integer~$N'_{\beta'}$ and an
  ordinal~$\lambda_{\beta'}$ such that, for any $n > N'_{\beta'}$,
  $\lambda_{n,\beta'} = \lambda_{\beta'}$.  We consider then two sub-cases
  depending on whether the ordinal~$\beta$ is a successor or a limit
  ordinal.

  \paragraph{Case B2a:} Let us suppose first that $\beta$ is a limit ordinal.
  For each ordinal $\beta' < \beta$, let us define the word~$u_{\beta'}$ by $u_{\beta'} =
  x\ropen{\lambda_{\beta'},\lambda_{\beta'+1}}$.  We claim that the sequence ${(u_{\beta'})}_{\beta'<\beta}$ is a
  prime factorization of~$x$.  It must be checked that $\sup \{ \lambda_{\beta'} \mid \beta'
  < \beta \}$ is equal to the length~$\gamma$ of~$x$.  But this follows from the
  equalities $\gamma = \sup \{\lambda_{n,\beta} \mid n < \omega \}$ and $\lambda_{n,\beta} = \sup \{ \lambda_{n,\beta'} \mid
  \beta' < \beta \}$ for each $n<\omega$.  Each word~$u_{\beta'}$ for $\beta' < \beta$ is prime
  since it occurs in the prime factorization of~$x_n$ for $n > N_\beta$.  For
  $n > N_\beta$, $\lambda_{n,\beta'+1}$ is equal to $\lambda_{\beta'+1}$.  The sequence
  ${(u_{\beta'})}_{\beta'<\beta}$ is densely non-increasing since each of its initial
  segments ${(u_{\beta'})}_{\beta'<\bar{\beta}}$ for $\bar{\beta} < \beta$ is densely
  non-increasing.

  \paragraph{Case B2b:} Let us now suppose that $\beta$ is a successor ordinal
  $\beta = \bar{\beta}+1$.  For $\beta' < \bar{\beta}$, let us define the word~$u_{\beta'}$ by
  $u_{\beta'} = x\ropen{\lambda_{\beta'},\lambda_{\beta'+1}}$.  Define also the word $u_{\bar{\beta}}$ by
  $u_{\bar{\beta}} = x\ropen{\gamma_{\bar{\beta}},\gamma}$ where $\gamma$ is the length of~$x$.  We
  claim that the sequence ${(u_{\beta'})}_{\beta'<\beta}$ is a prime factorization
  of~$x$.  As in the previous case, each word~$u_{\beta'}$ for $\beta' < \bar{\beta}$
  is prime since it occurs in the prime factorization of~$x_n$ for $n >
  N_{\beta'}$.  The last word~$u_{\bar{\beta}}$ is prime by
  Proposition~\ref{pro:limit} since each word $x_n\ropen{\lambda_{\bar{\beta}},\lambda_{n,\beta}}$ is
  prime for $n$ great enough.  The sequence ${(u_{\beta'})}_{\beta'<\bar{\beta}}$ without
  the last word~$u_{\bar{\beta}}$ is densely non-increasing since it is the
  prime factorization of $x_n\ropen{0,\lambda_{\bar{\beta}}}$ for $n > N_{\bar{\beta}}$.
\end{proof}

\subsection{Uniqueness of the factorization}\label{sec:uniqueness}

In this section, we prove that any word has at most one prime
factorization.  It is quite surprising that the uniqueness of the
factorization has been used in the proof of the existence. We start with a
technical lemma used for the proof of the crucial
Proposition~\ref{pro:longestprefix}.

\begin{lem}\label{lem:proddecprimes}
  Let $\bar{u} = {(u_\beta)}_{\beta<\alpha}$ be a non-increasing sequence of prime words.
  If $\alpha \ge 2$, the product $\prod_{\beta<\alpha}{u_\beta}$ is not prime.
\end{lem}
\begin{proof}
  Let $u$ be the product $\prod_{\beta<\alpha}{u_\beta}$.  If the
  sequence~$\bar{u}$ is constant, that is, if $u_\beta = u_0$ for any
  $\beta<\alpha$, the word $u = u_0^\alpha$ with $\alpha \ge 2$ is not
  primitive and thus it is not prime.

  Now suppose that the sequence~$\bar{u}$ is not constant.  If $\alpha$ is
  a successor ordinal $\alpha'+1$, the last word~$u_{\alpha'}$ of the
  sequence is a suffix of~$u$.  This suffix satisfies
  $u_{\alpha'} \ltlex u_0$ because $\bar{u}$ is non-increasing and not
  constant and since $u_0 \ltlex u$, it satisfies $u_{\alpha'} \ltlex u$.
  This proves that $u$ is not prime.

  Now suppose that $\alpha$ is a limit ordinal.  Suppose first that the
  sequence is ultimately constant in~$\alpha$.  There exists then some
  ordinal $\gamma < \alpha$ such that for any $\gamma < \beta < \alpha$,
  $u_\beta = u_\gamma$ holds.  The word $u_\gamma^{\alpha-\gamma}$ is a
  suffix of~$u$ and it satisfies $u_\gamma^{\alpha-\gamma} \ltlex u$.
  Indeed, one has $u_\gamma \ltlex u_0$ since $\bar{u}$ is non-decreasing,
  but not constant and thus $u_\gamma^{\alpha-\gamma} \ltlex u_0$ by
  Corollary~\ref{cor:ualphav}.  Combining this relation with $u_0 \ltlex u$
  since $u_0$ is a prefix of~$u$, yields
  $u_\gamma^{\alpha-\gamma} \ltlex u$. Therefore, the word~$u$ is not
  prime.

  Now suppose that the sequence is not ultimately constant.  For any
  ordinal $\gamma < \alpha$, there exist $\gamma < \beta < \beta' < \alpha$
  such that $u_\beta \gtlex u_{\beta'}$.  We claim that there is an
  ordinal~$\gamma$ such that $u_\gamma \ltstr u_0$.  Otherwise each
  word~$u_\beta$ satisfies $u_\beta \lepre u_0$.  Since each word~$u_\beta$
  is a prefix of the same word~$u_0$, the relation
  $u_\gamma \gtlex u_{\gamma'}$ implies $|u_\gamma| > |u_{\gamma'}|$.
  Since the lengths of the words~$u_\beta$ cannot strictly decrease
  infinitely often by the fundamental property of ordinals, there is a
  contradiction and there exists then an ordinal~$\gamma < \alpha$ such
  that $u_\gamma \ltstr u_0$. Then the suffix
  $v = \prod_{\gamma\le\beta<\alpha}{u_\beta}$ of~$u$ satisfies
  $v \ltlex u_0 \ltlex u$ and the word~$u$ is not prime.
\end{proof}

The following proposition is the key property used to establish the
uniqueness of the factorization in prime words.  It characterizes the first
prime word of the factorization as the longest prime prefix.  It extends
the classical result for finite words to transfinite words (see proof of~\cite[Thm~5.1.5]{Lothaire}).
\begin{prop}\label{pro:longestprefix}
  Let $\bar{u} = {(u_\beta)}_{\beta<\alpha}$ be a densely non-increasing
  sequence of prime words.  The word~$u_0$ is the longest prime prefix of
  the product $\prod_{\beta<\alpha}{u_\beta}$.
\end{prop}
\begin{proof}
  Let $u$ be the product $\prod_{\beta<\alpha}{u_\beta}$.  The word~$u_0$
  is clearly a prime prefix of~$u$.  It remains to prove that any
  prefix~$w$ of~$u$ such that $u_0 \ltpre w$ is not prime.

  We first suppose that the sequence ${(u_\beta)}_{\beta<\alpha}$ of prime
  words is non-increasing.  Let $x$ be a prefix of~$u$ such that
  $u_0 \ltpre x$.  This prefix~$x$ is equal to a product
  $(\prod_{\beta<\gamma}{u_\beta})u'$ where $1 \le \gamma < \alpha$ and
  $u'$ is a prefix of~$u_\gamma$ different from~$u_\gamma$.  If $u'$ is
  empty, then $\gamma \ge 2$, and the product
  $\prod_{\beta<\gamma}{u_\beta}$ cannot be prime by
  Lemma~\ref{lem:proddecprimes}.  If $u'$ is not empty, it is a suffix
  of~$x$.  Suppose that $x$ is prime.  One has $u_0 \ltlex x$ since
  $u_0 \ltpre x$, $x \lelex u'$ since $x$ is prime and $u'$ is a suffix
  of~$x$, $u' \ltlex u_\gamma$ since $u' \ltpre u_\gamma$ and
  $u_\gamma \lelex u_0$ since $\bar{u}$ is non-increasing.  Combining all
  these relations yields $u_0 \ltlex u_0$ and this is a contradiction.
  Therefore $x$ is not prime.

  We now suppose that the sequence ${(u_\beta)}_{\beta<\alpha}$ of prime
  words is not non-increasing. By Lemma~\ref{lem:decrprefix}, there exists
  a greatest ordinal~$\alpha'$ such that ${(u_\beta)}_{\beta<\alpha'}$ is
  non-increasing.  Furthermore, the ordinal~$\alpha'$ is limit and the
  sequence ${(u_\beta)}_{\beta<\alpha}$ is not ultimately constant
  in~$\alpha'$.

  If $x$ is a prefix of the product $u' = \prod_{\beta<\alpha'}{u_\beta}$,
  then the result follows from the previous case.  We now suppose that $u'$
  is a prefix of~$x$.  We claim that there exists an ordinal
  $\gamma < \alpha'$ such that $u_\gamma \ltstr u_0$.  Indeed, if
  $u_\beta \lepre u_0$ holds for any $\beta < \alpha'$, the length
  $|u_\beta|$ must decrease infinitely often before~$\alpha'$ since
  ${(u_\beta)}_{\beta<\alpha}$ is not ultimately constant in~$\alpha'$.  This
  is a contradiction with the fundamental property of ordinals.  There
  exists then some ordinal $\gamma < \alpha'$ such that
  $u_\gamma \ltstr u_0$.

  Since $u' \lepre x$, the suffix~$v$ of~$x$ such that
  $x = (\prod_{\beta<\gamma}{u_\beta})v$ satisfies $u_\gamma \ltpre v$.  It
  follows from $u_\gamma \ltstr u_0$ that $v \ltstr u_0 \ltlex x$ and the
  word $x$ is not prime.
\end{proof}

The next corollary uses the previous proposition to prove the uniqueness of
the factorization.
\begin{cor}\label{cor:uniquefact}
  For any word~$x$, there exists at most one densely non-increasing
  sequence ${(u_\beta)}_{\beta<\alpha}$ of prime words such that
  $x = \prod_{\beta<\alpha}{u_\beta}$.
\end{cor}
\begin{proof}
  Suppose there exist two distinct densely non-increasing sequences
  ${(u_\beta)}_{\beta<\alpha}$ and ${(u'_\beta)}_{\beta<\alpha'}$ such that
  $x = \prod_{\beta<\alpha}{u_\beta} = \prod_{\beta<\alpha'}{u'_\beta}$.
  Let $\gamma$ be the least ordinal such that $u_\gamma \neq u'_\gamma$.
  Let the ordinal~$\delta$ be equal to the sum
  $\prod_{\beta<\gamma}{|u_\beta|} = \prod_{\beta<\gamma}{|u'_\beta|}$.  By
  the previous proposition both $u_\gamma$ and $u'_\gamma$ are the longest
  prime prefix of the suffix $x\ropen{\delta,|x|}$ of~$x$ starting at
  position~$\delta$.  It follows that $u_\gamma = u'_\gamma$ and this is a
  contradiction.
\end{proof}

In the case of finite words, it can be shown~\cite[Prop.~5.1.6]{Lothaire}
that the last prime word of the prime factorization of a word~$x$ is the
least suffix (for the lexicographic ordering) of~$x$.  A similar result
does not hold for transfinite words.  Since the lexicographic ordering is
not well founded, a word may not have a least suffix.  Consider, for
instance, the $\omega$-word $x_0 = aba^2ba^3b \cdots$.  It does not have a
least suffix and its prime factorization
$x_0 = ab \cdot a^2b \cdot a^3b \cdots$ does not have a last factor.  Even
when the prime factorization of a word~$x$ has a last prime factor, the
word~$x$ may not have a least suffix.  Consider the $(\omega+1)$-word
$x_2 = x_0b$.  The prime factorization
$x_2 = ab \cdot a^2b \cdot a^3b \cdots b$ has a last factor~$b$, but this
word~$x_2$ does not have a least suffix.

Combining Corollary~\ref{cor:uniquefact} and
Proposition~\ref{pro:existsfact} gives Theorem~\ref{thm:main}.

\section*{Conclusion}

To conclude, let us sketch a few problems that are raised by our work.

In order to obtain a prime factorization for each transfinite word, we have
only required the sequence of prime words to be densely non-increasing.  It
seems interesting to characterize those words that have a decreasing
factorization.  We prove in Theorem~\ref{thm:factrat} that rational words
do have such a factorization, but they are not the only ones.  The
$\omega$-word $x = aba^2ba^3b \cdots$ has also the decreasing factorization
$x = ab \cdot a^2b \cdot a^3b \cdots$.

The algorithm given in the Appendix~\ref{sec:algo} outputs the
factorization of a rational word given by an expression~$e$ by inserting
markers in the duplicated expression~$\tau(e)$.  Even if the complexity of
this algorithm is polynomial in the size of~$\tau(e)$, the algorithm is
indeed exponential in the size of the expression~$e$. This is due to the
exponential blow up generated by the duplication. It could then be
interesting to design a better algorithm: this new algorithm could
determine which parts of the expression~$e$ have to be duplicated in order
to get a better complexity. In particular, it would be interesting to know
whether the exponential blow up is really needed.  Along the same lines, it
seems that it is possible to design an algorithm such that, given an
expression~$e$, it decides if the expression can be used to describe the
factorization of the corresponding rational word without any duplication.

We thank both referees for their constructive and helpful comments which
help us to improve the presentation of the paper.

\section*{Acknowledgements}

Carton is a member of the Laboratoire International Associ\'e SINFIN,
CONICET/Universidad de Buenos Aires--CNRS/Universit\'e de Paris and he is 
supported by the ECOS project PA17C04.  Carton is also partially funded by
the DeLTA project (ANR-16-CE40-0007). 

The authors are very grateful to the anonymous referees for reading the
first version of this paper with exceptional accurateness and for making
many suggestions for possible improvements.

\bibliographystyle{plain}
\bibliography{lyndon}

\appendix

\section{Rational words}\label{sec:rational}

The appendices are devoted to prove that, for a special kind of transfinite
words, the prime factorization can be effectively computed.  The result is
proved in Appendix~\ref{sec:algo} whence this section introduces these
special words called rational words. First some elementary properties of
their prime factorization are proved in Section~\ref{sec:factrat}. After
the introduction of the notion of cut in Section~\ref{sec:cuts} used to
define positions in a rational word, a description of any rational word by
a generalized finite automaton is presented in Sections~\ref{sec:automata}
and~\ref{sec:singleaut}. Then a last technical transformation, the
duplication operation, is defined in Section~\ref{sec:duplication}.  This
transformation is applied to the given expression before computing the
associated automaton and processing it with the algorithm presented in
Appendix~\ref{sec:algo}.  This algorithm is first described and an example
of its execution is presented in Section~\ref{sec:algodesc}. Before proving
the algorithm, some necessary auxiliary results are proved in
Section~\ref{sec:addprop}. Finally, five invariants are shown to hold in
Section~\ref{sec:invariants}, which then allow us to prove the correctness
of the algorithm and its complexity in Section~\ref{sec:algoproof}.

The class of \emph{rational words} is the smallest class of words that
contains the empty word~$\varepsilon$ and the letters and that is closed under
product and the iteration~$\omega$.  This means that each letter~$a$ is a
rational word and that if $u$ and~$v$ are two rational words, then both
words $uv$ and~$u^\omega$ are also rational.  A rational word is a word that can
be described by a rational expression using only concatenation and the $\omega$
operator.

All finite words are rational.  The word ${(a^\omega b^\omega b)}^\omega{(ab)}^\omega$ whose length
is $(\omega\cdot2 + 1)\cdot \omega + \omega = \omega^2+\omega$ is rational, but the $\omega$-word $aba^2ba^3 \cdots$
is not rational.  Notice that the length of a rational word is always less
than $\omega^\omega$.

\subsection{Factorization of rational words}\label{sec:factrat}

The following theorem states that the prime factorization of a rational
word has a very special form, namely it has a finite range made of rational
words.

Consider for instance the rational word $x = {(a^\omega b)}^\omega a^\omega$.  Its prime
factorization is $x = u_1^\omega u_2^\omega$ where $u_1 = a^\omega b$ and $u_2 = a$.
There are only two distinct prime factors and each of them is rational.

\begin{thm}\label{thm:factrat}
  For any rational word~$x$, there exists a finite decreasing sequence of
  rational prime words $u_1 \gtlex \cdots \gtlex u_n$ and ordinals $\alpha_1,\ldots,\alpha_n$
  less than $\omega^\omega$ such that $x = u_1^{\alpha_1} \cdots u_n^{\alpha_n}$.
\end{thm}

Let us make a few comments before proving the theorem.  Let $x$ be a
rational word and let ${(u_\beta)}_{\beta<\alpha}$ be its prime factorization.  The
previous theorem states first that the sequence ${(u_\beta)}_{\beta<\alpha}$ has a finite
range and is non-increasing.  Note that the second property is actually
implied by the first one by Corollary~\ref{cor:finiterange}.  The theorem
also states that each word occurring in ${(u_\beta)}_{\beta<\alpha}$ is also rational.
The fact that the exponents $\alpha_1,\ldots,\alpha_n$ are less than $\omega^\omega$ follows from
the fact that the length of each rational word is less than $\omega^\omega$.

In order to prove that the prime factorization of a rational word has
always the form given in Theorem~\ref{thm:factrat}, it is sufficient to
prove that this form is preserved by product and $\omega$-iteration.  The
preservation by product is already given by Lemma~\ref{lem:factproduct}.
The preservation by $\omega$-iteration is stated in Lemma~\ref{lem:factomega}
below.  The statement of this lemma is actually stronger than what is
really needed for the proof of Theorem~\ref{thm:factrat}, but this stronger
version is used later in the Appendix~\ref{sec:algo}.
Lemma~\ref{lem:productsuffix} is used to prove Lemma~\ref{lem:circularfact}
which is, in turn, used to prove Lemma~\ref{lem:factomega}.

\begin{lem}\label{lem:productsuffix}
  Given $n$ ordinal powers of prime words $u_1^{\alpha_1},\ldots,u_n^{\alpha_n}$ such that
  the product $u_1^{\alpha_1}\cdots u_n^{\alpha_n}$ is a power $v^\beta$ of a prime
  word~$v$, then either $v = u_n$ or $u_n^{\alpha_n}$ is a suffix of~$v$.
\end{lem}
\begin{proof}
  The proof is by induction on $n$.  If $n = 1$, $v = u_1$ and the result
  is obvious.  Now assume that $n > 1$.  If for each integer $1 \le i \le n-1$,
  $u_i \gtlex u_{i+1}$, then $v^\beta$ and $u_1^{\alpha_1}\cdots u_n^{\alpha_n}$ are two prime
  factorizations of the same word, which is impossible.  Hence, there
  exists an integer $1 \le i \le n-1$ such that $u_i \lelex u_{i+1}$.  By
  Lemma~\ref{lem:concatuv}, the word $u_i^{\alpha_i}u_{i+1}^{\alpha_{i+1}}$ is equal
  to~$w^\gamma$ for a prime word~$w$.  Moreover, by the same lemma, either $w =
  u_{i+1}$ and $\gamma \in \{\alpha_{i+1}, \alpha_i+\alpha_{i+1}\}$ or $w =
  u_i^{\alpha_i}u_{i+1}^{\alpha_{i+1}}$ and $\gamma = 1$.  If $i \le n-2$, then the
  induction hypothesis gives obviously the result.  If $i = n-1$, then
  $w^\gamma$ is equal to $u_{n-1}^{\alpha_{n-1}}u_n^{\alpha_n}$ with either $w = u_n$ and
  $\gamma \in \{\alpha_n,\alpha_{n-1}+\alpha_n\}$ or $w = u_{n-1}^{\alpha_{n-1}}u_n^{\alpha_n}$ and $\gamma =
  1$.  On the other hand, the hypothesis can be written $u_1^{\alpha_1} \cdots
  u_{n-2}^{\alpha_{n-2}}w^\gamma = v^\beta$.  By the induction hypothesis, either $v = w$
  or $w^\gamma$ is a suffix of~$v$.  This gives rise to four cases that we
  consider.

  If $w = u_n$ and $v = w$, then $v = u_n$ trivially.  If $w = u_n$ and
  $w^\gamma$ is a suffix of~$v$, the ordinal~$\gamma$ is either $\alpha_n$ or
  $\alpha_{n-1}+\alpha_n$.  Therefore $u_n^{\alpha_n}$ is a suffix of~$v$.  If $w =
  u_{n-1}^{\alpha_{n-1}}u_n^{\alpha_n}$ and $v = w$, then $u_n^{\alpha_n}$ is a suffix
  of~$v$ trivially.  Finally if $w = u_{n-1}^{\alpha_{n-1}}u_n^{\alpha_n}$ and $w^\gamma$
  is a suffix of~$v$, the ordinal $\gamma$ is then $1$ and $u_n^{\alpha_n}$ is a
  suffix of~$v$.
\end{proof}

\begin{lem}\label{lem:circularfact}
  Given $n$ ordinal powers of prime words $u_1^{\alpha_1},\ldots,u_n^{\alpha_n}$ there
  exists an integer $1 \le k \le n$, a prime word $v$~and an ordinal~$\beta$ such
  that $v^\beta = u_{k+1}^{\alpha_{k+1}} \cdots u_n^{\alpha_n}u_1^{\alpha_1} \cdots u_k^{\alpha_k}$ and $v
  \lelex u_k$.  Furthermore, if each $u_i$ is rational and each $\alpha_i <
  \omega^\omega$, then $v$ is also rational and $\beta < \omega^\omega$.
\end{lem}
\begin{proof}
  We first prove by induction on~$n$ that there exist an integer $1 \le k \le
  n$, a prime word $v$~and an ordinal~$\beta$ such that $v^\beta =
  u_{k+1}^{\alpha_{k+1}} \cdots u_n^{\alpha_n}u_1^{\alpha_1} \cdots u_k^{\alpha_k}$.  If $n = 1$, the
  result is clear with $v = u_1$ and $\beta = \alpha_1$.

  Now let $n > 1$. If $u_1 = \cdots = u_n$, it suffices to take $v = u_1$ and $\beta
  = \alpha_1 + \cdots + \alpha_n$. Otherwise, there exist an integer $1 \le i \le n$ such that
  $u_i \ltlex u_{i+1}$ where $n+1$ should be understood as~$1$.  By
  Lemma~\ref{lem:concatuv}, the word $u_i^{\alpha_i}u_{i+1}^{\alpha_{i+1}}$ is equal
  to $w^\gamma$ where $w$ is a prime word.  The induction hypothesis is now
  applied to $u_1^{\alpha_1},\ldots,u_{i-1}^{\alpha_{i-1}},w^{\gamma} ,u_{i+1}^{\alpha_{i+1}}
  ,\ldots,u_n^{\alpha_n}$ if $i < n$ and to $w^{\gamma},u_{2}^{\alpha_{2}},\ldots,u_{n-1}^{\alpha_{n-1}}$
  if $i = n$.

  We now prove the second part, namely that the word~$v$ satisfies $v
  \lelex u_k$.  This is a direct consequence of
  Lemma~\ref{lem:productsuffix}: assume, by contradiction, that $u_k \ltlex
  v$.  By Corollary~\ref{cor:ualphav}, $u_k^{\alpha_k} \ltlex v$ which is
  impossible since $v$ is prime.

  The fact that $v$ is rational and $\beta < \omega^\omega$ under the given assumptions
  is obvious from the constructions of $v$ and~$\beta$.
\end{proof}

\begin{lem}\label{lem:factomega}
  Let $x = u_1^{\alpha_1} \cdots u_n^{\alpha_n}$ be the prime factorization of the
  word~$x$.  There exist an integer $1 \le k \le n-1$ ($k = 1$ if $n = 1$) such
  that the prime factorization of $x^\omega$ is either $u_1^{\alpha_1} \cdots
  u_k^{\alpha_k}v^{\beta\omega}$ or $u_1^{\alpha_1} \cdots u_{k-1}^{\alpha_{k-1}}v^{\alpha_k+\beta\omega}$
  ($u_1^{\alpha_1\omega}$ if $n = 1$) where $v^\beta = u_{k+1}^{\alpha_{k+1}} \cdots
  u_n^{\alpha_n}u_1^{\alpha_1} \cdots u_k^{\alpha_k}$.  Furthermore, if each $u_i$ is rational
  and each $\alpha_i < \omega^\omega$, then $v$ is also rational and $\beta < \omega^\omega$.
\end{lem}

\begin{proof}
  The result for $n = 1$ is obvious. We now assume that $n \ge 2$.  We apply
  Lemma~\ref{lem:circularfact} to the sequence $u_1^{\alpha_1},\ldots, u_n^{\alpha_n}$ to
  obtain an integer~$1 \le k \le n$, an ordinal~$\beta$, and a prime word~$v$
  such that $v^\beta = u_{k+1}^{\alpha_{k+1}} \cdots u_n^{\alpha_n}u_1^{\alpha_1} \cdots u_k^{\alpha_k}$.
  The case $k = n$ would give two prime factorizations $u_1^{\alpha_1} \cdots
  u_n^{\alpha_n}$ and $v^\beta$ of the word~$x$, which is impossible since $n \ge 2$.
  By Lemma~\ref{lem:circularfact}, the word~$v$ satisfies $v \lelex u_k$.

  If $v \ltlex u_k$, then $u_1^{\alpha_1} \cdots u_k^{\alpha_k}v^{\beta\omega}$ is indeed the prime
  factorization of~$x^\omega$.  If $v = u_k$, then $u_1^{\alpha_1} \cdots
  u_{k-1}^{\alpha_{k-1}}v^{\alpha_k+\beta\omega}$ is the prime factorization of~$x^\omega$.

  The fact that $v$ is rational and $\beta < \omega^\omega$ under the given assumptions
  follows from Lemma~\ref{lem:circularfact}.
\end{proof}

By a similar argument, it could be proved that the prime factorization
of~$x^m$ for an integer~$m$ has either the form $x^m = u_1^{\alpha_1} \cdots
u_k^{\alpha_k} v^{\beta(m-1)} u_{k+1}^{\alpha_{k+1}} \cdots u_n^{\alpha_n}$ or the form $x^m =
u_1^{\alpha_1} \cdots u_{k-1}^{\alpha_{k-1}} v^{\alpha_k+\beta(m-1)} u_{k+1}^{\alpha_{k+1}} \cdots
u_n^{\alpha_n}$.

We now come to the proof of Theorem~\ref{thm:factrat}.
\begin{proof}[Proof of Theorem~\ref{thm:factrat}]
  Each word of length~$1$ is prime.  It suffices then to prove that if the
  rational words $x$ and~$y$ have a prime factorization of the required
  form, then the words $xy$ and $x^\omega$ also have a prime factorization of
  the required form.  The result for~$xy$ follows from
  Lemma~\ref{lem:finitefact} and the result for~$x^\omega$ follows from
  Lemma~\ref{lem:factomega}.
\end{proof}

The following lemma is used in Section~\ref{sec:invariants} to prove one
invariant of the algorithm.
\begin{lem}\label{lem:squareend}
  Let $x$ and $y$ be two words.  If the word $xy^2$ is prime, then the word
  $xy^\omega$ is also prime.
\end{lem}
\begin{proof}
  Let $u$ be the prime word $xy^2$.  The word $xy^\omega$ is equal to $uy^\omega$.
  We first verify that each suffix~$z$ of~$uy^\omega$ satisfies $uy^\omega \lelex z$.
  Such a suffix is either of the form $x'y^\omega$ where $x'$ is a suffix of~$x$
  or of the form $y'y^\omega$ where $y'$ is a suffix of~$y$.  If $z$ is equal to
  $x'y^\omega$, then $x'y^2$ is a suffix of~$u$.  Since $u$ is prime, then
  either $u \ltstr x'y^2$ or $u = x'y^2$ holds.  If $u \ltstr x'y^2$, then
  $uy^\omega \ltstr x'y^\omega$ and if $u = x'y^2$, then $uy^\omega = x'y^\omega$.  Thus, in
  any case, $uy^\omega \lelex z$.  If $z$ is equal to $y'y^\omega$, then $y'y$ is a
  suffix of $u$.  Since $u$ is prime, then either $u \ltstr y'y$ or $u =
  y'y$ holds.  If $u \ltstr y'y$, then $uy^\omega \ltstr y'y^\omega$ and if $u =
  y'y$, then $uy^\omega = y'y^\omega$.

  It remains to show that $xy^\omega$ is primitive.  If $xy^\omega$ is not primitive,
  it is equal, by Lemma~\ref{lem:power}, to $z^\alpha$ for some word~$z$ and
  some limit ordinal~$\alpha$.  We first claim that $xy^2$ is a prefix of~$z$.
  If $xy^2$ is not prefix of~$z$, there exist two words $z_1$ and~$z_2$ and
  two ordinals $\alpha_1$ and~$\alpha_2$ such that $z = z_1z_2$ and
  $\alpha = \alpha_1 + 1 + \alpha_2$ and $xy^2 = z^{\alpha_1}z_1$ and $y^\omega = z_2z^{\alpha_2}$.  If
  $\alpha_1 \ge 1$, the word~$z_1$ is a suffix of~$u = xy^2$ and a proper prefix
  of~$u$.  This contradicts the fact that $u$ is prime.  The word~$z_1$ is
  thus equal to~$u$, and this proves the claim that $xy^2$ is a prefix
  of~$z$.  Note that $|z_1| \ge |y| \cdot 2$ since $z_1 = xy^2$.  Since
  $y^\omega = z_2z^{\alpha_2}$, the first occurrence of~$z$ in $y^\omega$ has a prefix of
  the form $y'y$ where $y'$ is a suffix of~$y$.  If follows that $y'y$ is
  also a prefix of~$u$ since $u = z_1$.  This contradicts again the fact
  that $u$ is prime.
\end{proof}

\subsection{Cuts}\label{sec:cuts}

We now introduce the notion of a cut of a word.  This notion is used to
describe the prime factorization of a word. A \emph{cut} of a word~$x$ is a
factorization $x = x_1x_2$ into two factors.  It is merely denoted by a dot
between the two factors as in $x = x_1 \cdot x_2$.  The trivial cuts are the
two factorizations $x = \varepsilon \cdot x$ and $x = x \cdot \varepsilon$ where one of the two factors
is empty.  Since each factorization is characterized by the length of the
prefix~$x_1$, the cuts of~$x$ can be identified with ordinals between $0$
and $|x|$.  The trivial cuts correspond to the ordinals $0$ and~$|x|$.  For
instance, consider again the word $x = {(a^\omega b)}^\omega a^\omega$.  The cut $x = {(a^\omega
b)}^3 a^\omega \cdot b {(a^\omega b)}^\omega a^\omega$ corresponds to the ordinal $(\omega + 1) 3 + \omega =
\omega\cdot4$.

Given the prime factorization $x = u_1^{\alpha_1} \cdots u_n^{\alpha_n}$ of
a rational word, we introduce two kinds of particular cuts of~$x$.
Intuitively, main cuts are between two different prime factors and
secondary cuts are between two occurrences of the same prime factor.
Formally, each factorization $x = x_1 \cdot x_2$ with
$x_1 = u_1^{\alpha_1} \cdots u_k^{\alpha_k}$ and
$x_2 = u_{k+1}^{\alpha_{k+1}} \cdots u_n^{\alpha_n}$ for some
$1 \le k \le n-1$ is called a \emph{main} cut of~$x$.  By convention, the
two trivial factorizations $\varepsilon \cdot x$ and $x \cdot \varepsilon$
are considered as main cuts.  A factorization $x = x_1 \cdot x_2$ with
$x_1 = u_1^{\alpha_1} \cdots u_k^{\beta_1}$ and
$x_2 = u_k^{\beta_2}u_{k+1}^{\alpha_{k+1}} \cdots u_n^{\alpha_n}$ where
$\alpha_k = \beta_1 + \beta_2$ and $\beta_1,\beta_2 \neq 0$ is called a
\emph{secondary} cut of~$x$.

We illustrate the notion of cuts by two examples.  Let $x$ be the word
${(bba)}^\omega a$.  Its prime factorization is $x = u_1^2 u_2^\omega u_3$
where $u_1 = b$, $u_2 = abb$ and $u_3 = a$.  Hence, the two factorizations
$x = b^2 \cdot {(abb)}^\omega a$ and $b^2{(abb)}^\omega \cdot a$ are main cuts.
The two factorizations $x = b \cdot b{(abb)}^\omega a$ and
$x = b^2{(abb)}^3 \cdot {(abb)}^\omega a$ are secondary cuts.  Note that the
cut $x = {(bba)}^2b \cdot b{(abb)}^\omega a$ is neither main nor secondary.

The next example is used later to illustrate the algorithm.  The prime
factorization of ${(a^\omega b)}^\omega a^\omega$ is $u_1^\omega u_2^\omega$
where $u_1 = a^\omega b$ and $u_2 = a$.  Hence, the factorization
$x = {(a^\omega b)}^\omega \cdot a^\omega$ is a main cut.  The factorization
$x = {(a^\omega b)}^3 \cdot {(a^\omega b)}^\omega a^\omega$ is a secondary cut.
Note that the cut
$x = {(a^\omega b)}^3 a^\omega \cdot b {(a^\omega b)}^\omega a^\omega$ is
neither main nor secondary.

We can now rephrase Lemmas~\ref{lem:factproduct} and~\ref{lem:factomega}
in terms of cuts.
\begin{cor}\label{cor:cutsproduct}
  Given two rational words $x$ and~$y$, the main cuts of $xy$ are
  main cuts of $x$ or~$y$.  Secondary cuts of~$xy$ are main or secondary
  cuts of $x$ or~$y$.
\end{cor}

The statement of the previous corollary means that if $u\cdot v$ is a main
of~$xy$ then there exists a main cut $u\cdot v'$ of~$x$ with $v = v'y$ or
there exists a main cut $u'\cdot v$ of~$y$ with $u = xu'$.  Note that some
main cuts of~$x$ and~$y$ may not give rise to cuts of~$xy$.  A similar
comment could be made after the following corollary.

\begin{cor}\label{cor:cutsomega}

  Given a rational word~$y$, the main cuts of $y^\omega$ are main cuts
  of~$y$.  Furthermore, all occur within the prefix~$y$. Secondary cuts
  of~$y^\omega$ are main or secondary cuts of~$y$.
\end{cor}

\subsection{Automata}\label{sec:automata}

We introduce automata accepting transfinite words which generalize usual
automata accepting finite words.  In the next section, we consider such
automata that accept a single transfinite word. It turns out that the
accepted word is then a rational word.  More precisely, a transfinite word
is rational if and only if there exists an automaton accepting this word
and no other word.  The automaton is then a finite representation of the
rational transfinite word.  We design an algorithm computing the prime
factorization working on the automaton accepting the rational word.

B\"uchi automata~\cite{Buchi65} on transfinite words are a generalization
of usual (Kleene) automata on finite words, with an additional special
transition function for limit ordinals.  States reached at limit points
depend only on states reached before.

An \emph{automaton} $\mathcal{A}$ is a 5-tuple $(Q,A,E,I,F)$ where $Q$ is the finite
set of states, $A$ the finite alphabet, $E \subseteq (Q \times A \times Q)\cup(2^Q\times Q)$ the set
of transitions, $I \subseteq Q$ the set of initial states and $F \subseteq Q$ the set of
final states.

Transitions are either of the form $(q,a,q')$ or of the form $(P,q)$ where
$P$ is a subset of~$Q$.  A transition of the former case is called a
\emph{successor transition} and it is denoted by $p \trans{a} q$.  These
are the usual transitions.  A transition of the latter case is called a
\emph{limit transition} and it is denoted by $P \rightarrow q$.  These are the
additional transitions.  All automata that we consider are deterministic:
for each pair $(p,a)$ in $Q \times A$, there exists at most one state~$q$ such
that $p \trans{a} q$ is a successor transition and for each subset $P \subseteq Q$,
there exists at most one state~$q$ such that $P \rightarrow q$ is a limit transition.

We now explain how these automata accept transfinite words. Before
describing a run in an automaton, we define the cofinal set of a sequence
at some limit point.

Let $c = {(q_\gamma)}_{\gamma<\alpha}$ be an $\alpha$-sequence of states and let $\beta\le\alpha$ be a limit
ordinal.  The \emph{limit set} of~$c$ at~$\beta$ is the set of states that
occur cofinally before the limit ordinal~$\beta$.  It is formally defined as
follows.
\begin{displaymath}
  \lim({(q_\gamma)}_{\gamma<\beta}) = \{ q \in Q \mid \forall \;\; \beta' < \beta
  \quad
  \exists \gamma \;\; \beta' < \gamma < \beta \land q = q_\gamma \}
\end{displaymath}

Let $\mathcal{A} = (Q,A,E,I,F)$ be an automaton. A \emph{run} $c$ labeled by a word
$x = {(a_\gamma)}_{\gamma<\alpha}$ of length~$\alpha$ from $p$ to $q$ in~$\mathcal{A}$ is an
$(\alpha+1)$-sequence of states $c = {(q_\gamma)}_{\gamma\le\alpha}$ such that:
\begin{enumerate} \itemsep=0cm
\item $q_0=p$ and $q_\alpha=q$;
\item for any ordinal $\beta < \alpha$, $q_\beta \trans{a_\beta} q_{\beta+1}$
  is a successor transition of~$\mathcal{A}$;
\item for any limit ordinal $\beta \le \alpha$, $\lim({(q_\gamma)}_{\gamma<\beta}) \rightarrow q_\beta$ is a
  limit transition of~$\mathcal{A}$.
\end{enumerate}

\noindent
The word $x = {(a_\gamma)}_{\gamma<\alpha}$ is called the \emph{label} of the run~$c$. The
run is \emph{successful} if and only if $p$ is initial ($p \in I$) and $q$
is final ($q \in F$).  A word is \emph{accepted} by the automaton if and only
if it is the label of a successful run.  As already mentioned, the cuts
of a word~$x$ can be identified with the ordinals between $0$ and~$|x|$.
Therefore, a run maps each cut to a state.  We illustrate the definition
of a run with the following example.




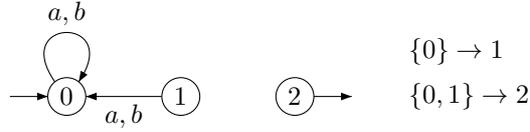
\begin{figure}[htbp]
  \begin{center}
    \gasset{Nw=5,Nh=5,loopdiam=6}
    \begin{picture}(60,15)(-5,-3)
    \node[Nmarks=i](R0)(0,0){$0$}
    \node(R1)(15,0){$1$}
    \node[Nmarks=f](R2)(30,0){$2$}
    \drawloop(R0){$a,b$}
    \drawedge(R1,R0){$a,b$}
    \put(45,6){\makebox(0,0)[l]{$\{0\} \rightarrow 1$}}
    \put(45,0){\makebox(0,0)[l]{$\{0,1\} \rightarrow 2$}}
   \end{picture}
    \caption{Automaton accepting words of length~$\omega^2$}%
    \label{fig:automaton1}
  \end{center}
\end{figure}

\begin{exa}
  The deterministic automaton pictured in Figure~\ref{fig:automaton1}
  accepts words of length~$\omega^2$.  Indeed let $u$ be a $\omega^2$-word
  ${(c_\beta)}_{\beta<\omega^2}$ where $c_\beta \in \{a, b\}$.  A run accepting $u$ is the
  $\omega^2+1$-sequence ${(q_\beta)}_{\beta\le\omega^2}$ where $q_\beta = 0$ if $\beta = \omega\cdot k_1 + k_0$
  with $k_0 \neq 0$ or $k_1 = 0$, $q_\beta = 1$ if $\beta = \omega \cdot k_1$ with $k_1 \neq 0$
  and $q_{\omega^2} = 2$.
\end{exa}




As usual, a \emph{loop} in such an automaton is a run from a state~$q$ to
the same state~$q$.

\subsection{Automata for a single word}\label{sec:singleaut}

For a given rational expression~$e$ denoting a single word~$x$, we describe
the construction of an automaton~$\mathcal{A}_e$ which accepts $x$ but no other word.
This automaton depends on the expression~$e$.  Two different expressions
denoting the same word may yield different automata.  We describe the
construction on an example.  The general case is straightforward.

Consider the rational expression ${(a^\omega b)}^\omega a^\omega$.  It is first flatten to
give the word $(a\omega b)\omega a\omega$ over the extended alphabet $A \cup \{(,),\omega\}$.  Let
$n$ be the number occurrences of letters in $A \cup \{\omega\}$ in this word.  In
our example, this number~$n$ is equal to~$6$.  The integers from~$0$ to
$n-1$ are then inserted before the letters in $A \cup \{\omega\}$ and the
integer~$n$ is added at the end of the word to obtain the word
$(0a1\omega2b)3\omega4a5\omega6$ over the alphabet $A \cup \{(,),\omega\} \cup \{0,1,\ldots,n\}$.

This allows to directly get an automaton in the following way.  The
integers from $0$ to~$n$ are its states.  The state~$0$ is the initial one
and $n$ is the unique final state. We now describe its successor and limit
transitions.

There is no transition from state~$n$.  Each integer $0 \le i \le n-1$ has been
inserted before a letter which is either a letter $a \in A$ or~$\omega$.  If $i$
lies just before a letter $a \in A$, there is a successor transition $i
\trans{a} (i+1)$.  If $i$ lies before a symbol~$\omega$, there are a successor
transition from~$i$ and a limit transition defined as follows.  Let $j-1$
be the integer just before the first letter of the sub-expression under
this~$\omega$ and let this letter be~$a$. Note that $j$ satisfies $j \le i$.  The
successor transition is then the transition $i \trans{a} j$ and the limit
transition is $\{j,j+1,\ldots,i\} \rightarrow (i+1)$.  Note that both transitions $(j-1)
\trans{a} j$ and $i \trans{a} j$ enter the same state $j$ and have the same
label.

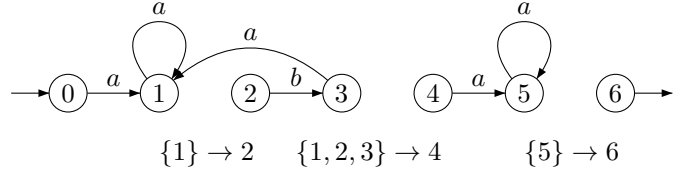
\begin{figure}[htbp]
  \begin{center}
    \gasset{Nw=5,Nh=5,loopdiam=7}
    \begin{picture}(72,25)(0,-8)
    \node[Nmarks=i](0)(0,0){$0$}
    \node(1)(12,0){$1$}
    \node(2)(24,0){$2$}
    \node(3)(36,0){$3$}
    \node(4)(48,0){$4$}
    \node(5)(60,0){$5$}
    \node[Nmarks=f](6)(72,0){$6$}
    \drawedge(0,1){$a$}
    \drawloop(1){$a$}
    \drawedge(2,3){$b$}
    \drawedge[ELside=r,curvedepth=-6](3,1){$a$}
    \drawedge(4,5){$a$}
    \drawloop(5){$a$}
    \put(12,-8){\makebox(0,0)[l]{$\{1\} \rightarrow 2$}}
    \put(30,-8){\makebox(0,0)[l]{$\{1,2,3\} \rightarrow 4$}}
    \put(60,-8){\makebox(0,0)[l]{$\{5\} \rightarrow 6$}}
    \end{picture}
    \caption{Automaton for ${(a^\omega b)}^\omega a^\omega$}%
    \label{fig:autom-aoboao}
  \end{center}
\end{figure}

Applying this construction to the expression ${(a^\omega b)}^\omega a^\omega$ gives the
automaton pictured in Figure~\ref{fig:autom-aoboao}.

The automata constructed by the algorithm given above have special
properties that we now give.  It has $n+1$ states, namely $\{0,\ldots,n\}$ where
$n$ is the total number of letters in~$A$ and~$\omega$s in the expression.
\begin{itemize}
\item The initial state is $0$ and the unique final state is $n$.
\item There is no successor transition leaving state~$n$ and for any state
  $0 \le i \le n-1$, there is exactly one transition $i \trans{a} j$
  leaving~$i$. This unique state~$j$ is denoted, as usual, by~$i \cdot a$.
  Furthermore the state~$j$ satisfies $j \le i+1$.  A transition $i \trans{a}
  (i+1)$ is called a \emph{forwards} transition and a transition $i
  \trans{a} j$ where $j \le i$ is called a \emph{backwards} transition.  For
  any backwards transition $i \trans{a} j$, there exists also a transition
  $(j-1) \trans{a} j$.  Therefore all transitions entering a given state
  have the same label.
\item Each limit transition has the form $\{j,j+1,\ldots,i\} \rightarrow i+1$ where $1 \le j
  \le i \le n-1$ and there exists such a transition if and only if there exists
  a backwards transition $i \trans{a} j$.
\item For any two limit transitions $P \rightarrow (i+1)$ and $P' \rightarrow (i'+1)$ where $P
  = \{j,\ldots,i\}$ and $P' = \{j',\ldots,i'\}$, then either $P$ and~$P'$ are
  disjoint, that is, $P \cap P' = \emptyset$, or one is contained in the other, that is,
  $P \subseteq P'$ or $P' \subseteq P$.  This means that the cycles are well-nested.
\item If there is a limit transition $P \rightarrow (i+1)$, there is neither another
  limit transition $P' \rightarrow (i+1)$ nor a successor transition $j \rightarrow (i+1)$.
\item Given a state~$i$, and two runs $\rho$ and~$\rho'$ starting from~$i$,
  either $\rho$ is a prefix of~$\rho'$ or $\rho'$ is a prefix of~$\rho$.  This is due
  to the strong determinism of the automaton: there is a single successor
  transition from any state and for any subset~$P$, there is at most one
  limit transition of the form $P \rightarrow (i+1)$.
\item Given a state~$i$, either it is reached by successor transitions or
  by a single limit transition but never by both types of transitions.  In
  the former case, all successor transitions have the same label.
\end{itemize}

\noindent
The last remark is derived by looking at the letter preceding state~$i$.
This letter is either a letter~$a$ or~$\omega$.  In the former case, all
transitions reaching~$i$ are successor and labeled by the letter~$a$.  In
the latter, it is reached by a unique limit transition.  Obviously only one
of these two possibilities may happen. Moreover, the initial state~$0$ is
the only state that is not reached by any transition.

Recall that a run of an automaton~$\mathcal{A}$ labelled by a (transfinite)
word~$x$ is the sequence of states visited by the automaton while
processing~$x$. For instance, on the word $x = {(a^\omega b)}^\omega a^\omega$, the run~$\rho$
of the automata given Figure~\ref{fig:autom-aoboao} is $0{(1^\omega 23)}^\omega 45^\omega
6$. In such a run, as soon as the automaton visits twice the same state,
there is a loop.  The \emph{entry state} of a loop is the state that
is accessible from the initial state~$0$ by the shortest run.  This is
well-defined since any two runs from~$0$ are prefix of each other.  This is
also the first state of the loop that is visited by the run from~$0$ to
the final state.  Due to the way the states of the automaton are numbered,
the entry state of a loop is always the smallest state that belongs to the
cycle.

The rest of the section is devoted to properties of the automaton
$\mathcal{A}_e$ obtained from a rational expression~$e$.  Such an
expression~$e$ is assumed to be fixed, and the automaton~$\mathcal{A}_e$ is
merely denoted by~$\mathcal{A}$ until the end of the section.  We start
with a first property of loops in~$\mathcal{A}$.

\begin{lem}\label{lem:loop}
  Let $P$ be a loop of~$\mathcal{A}$ and let $p_0$ be its entry state. Then
  the label of~$P$ from~$p_0$ to~$p_0$ has a last letter~$a$ and can be
  factorized as~$ya$.  The word~$x$ accepted by~$\mathcal{A}$ can be
  factorized as $x = x_1a{(ya)}^\omega x_2$ where the run of~$\mathcal{A}$
  on~$x$ reaches $p_0$ after~$x_1a$.
\end{lem}
\begin{proof}
  Consider a loop $P$ in which the set of visited states is
  $\{p_0, p_1, \ldots, p_k\}$.  Note that this set coincides with
  $\{p_0, p_0+1, \ldots, p_0+k\}$.  Moreover, the greatest state
  $p_k = p_0+k$ goes back to~$p_0$ using a successor transition labeled by
  a letter~$a$. The label of the loop is then a word of the form~$ya$ for
  some word~$y$.  Then due to the way the automaton~$\mathcal{A}$ is built,
  the state~$p_0$ is reached from the state $p_0-1$ by the successor
  transition $(p_0-1) \trans{a} p_0$ which proves the result.
\end{proof}

Given the automaton~$\mathcal{A}$, we now introduce two families of
automata ${_i\mathcal{A}_j}$ and ${_i\mathcal{A}_j^\#}$ built from
$\mathcal{A}$.  Let $0 \le i < j \le n$ be two states such that there
exists no backwards transition $k \trans{a} k'$ with $k' \le i \le k$.
This means that $i$ is not contained in any loop of~$\mathcal{A}$.  We then
build two new automata ${_i\mathcal{A}_j}$ and ${_i\mathcal{A}_j^\#}$.  The
latter one is obtained from the former one.

The automaton~${_i\mathcal{A}_j}$ is obtained by erasing the unique
successor transition leaving $j$ if $j < n$. In this automaton, $i$ is the
initial state and $j$ is the unique final state; it accepts a unique
transfinite rational word denoted by~${_{i}x_j}$.  The
automaton~${_i\mathcal{A}_j}$ only uses states in the interval $[i,j]$.
Indeed, it does not use any state $k' < i$ since there is no backwards
transition $k \trans{a} k'$ with $k' \le i \le k$.  It does not use any
state $k' > j$ since the successor transition leaving~$j$ has been removed.
Note that $\mathcal{A} = {_0\mathcal{A}_n}$.

We now come to ${_i\mathcal{A}_j^\#}$.  It is built by adding to
${_i\mathcal{A}_j}$ two transitions: a successor one and a limit one.  The
additional successor transition is $j \trans{a} (i+1)$ where $a$ is the
label of the transition $i \trans{a} (i+1)$.  The additional limit
transition is $\{i+1,\ldots,j\} \rightarrow j$.  The new automaton is of a
new type: it does not accept a single word because there is now a successor
transition leaving the final state. It accepts actually all powers of
${_{i}x_j}$, that is the set
${_{i}x_j^\#} = \{ {_{i}x_j^\alpha} \mid\text{$\alpha$ ordinal} \}$.  Such an
automaton is used in parallel with the automaton~${_i\mathcal{A}_n}$ to
detect powers of a prime word.  An example of an automaton
${_i\mathcal{A}_j^\#}$ is pictured in Figure~\ref{fig:autom-0a4} below.

In order to detect powers of~${_{i}x_j}$, we consider the product automaton
${_i\mathcal{A}_n} \times {_i\mathcal{A}_j^\#}$.  As in
${_i\mathcal{A}_n}$, two runs starting in the same state are prefix of each
other.  Therefore, there is a unique maximal run in
${_i\mathcal{A}_n} \times {_i\mathcal{A}_j^\#}$.  It is labelled by the
longest prefix of~${_{i}x_n}$ of the form ${(_{i}x_j)}^\beta y$ where $\beta$ is
an ordinal and $y$ is a prefix of ${_{i}x_j}$. The algorithm uses this
maximal run.

Now consider a loop in ${_i\mathcal{A}_n} \times {_i\mathcal{A}_j^\#}$.  The
entry state of such a loop is then defined as the state of the loop that is
the first one reached in the maximal run.  The strong determinism of the
first component ensures that, if a state $(q,q')$ is reached from $(k,k')$
by a forward successor transition, in the run, each occurrence of the state
$(q,q')$ occurs just after an occurrence of the state $(k,k')$. Similarly,
if the state $(k,k')$ is in a loop, in the run, the entry state of the loop
occurs before each occurrence of $(k,k')$.

It is easy to check that an entry state $(p,p')$ of a loop in
${_i\mathcal{A}_n} \times {_i\mathcal{A}_j^\#}$ satisfies that $p$ or $p'$
is the entry state of a loop in the corresponding automaton. Indeed, to
have a state $(q,q')$ belonging to a loop of
${_i\mathcal A}_n \times {_i\mathcal{A}_j^\#}$, it is needed that $q$ and
$q'$ do belong to loops of each component. The automaton reaches such a
state for the first time when, for the first time, this condition on the
two states is satisfied. This implies that one or the other is an entry
state. Exactly as it happened for loops of ${_0\mathcal{A}_n}$, given the
entry state $(p,p')$ of a loop of
${_i\mathcal{A}_n} \times {_i\mathcal{A}_j^\#}$, the loop can be described
by the finite sequence of states visited in the loop. The label of the loop
is a transfinite word. This word is a finite power of the word labeling the
loop over $p$ in ${_i\mathcal{A}_n}$, as well as of the word labeling the
loop over $p'$ in ${_i\mathcal{A}_j^\#}$.

We now show that the automaton ${_i\mathcal{A}_n} \times {_i\mathcal{A}_j^\#}$
satisfies that a state either is reached by a successor transition or is
reached by a limit one, but never by both. Observe first that
${_i\mathcal{A}_n}$ does satisfy this property. Then if the state
$(r,r')$ is reached by a successor transition, $r$ is reached by such a
transition in ${_i\mathcal{A}_n}$ and if the state $(r,r')$ is reached by a
limit transition, $r$ is reached by such a transition in
${_i\mathcal{A}_n}$. As for $r$ it cannot be that both cases happen, we get
the desired result. Using the same argument, we show that if a state
$(r,r')$ is reached by a limit transition, it cannot be reached by any
successor transition.

\begin{figure}[htbp]
  \begin{center}
    \gasset{Nw=5,Nh=5,loopdiam=7}
    \begin{picture}(15,30)(0,-8)
    \node[Nmarks=i](0)(0,0){$0$}
    \node[Nmarks=f](1)(15,0){$1$}
    \drawedge(0,1){$a$}
    \drawloop(1){$a$}
    \put(2,-8){\makebox(0,0)[l]{$\{1\} \rightarrow 1$}}
    \end{picture}
    \caption{The special automaton for $a^\#$}%
    \label{fig:autom-asharp}
  \end{center}
\end{figure}
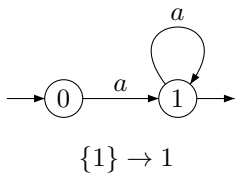

We now state three lemmas.  The first one is used in the proof of the
second one and the two last ones are used to prove the correctness of the
algorithm.

\begin{lem}\label{lem:entryproduct}
  An entry state of a loop of the automaton ${_i\mathcal{A}_n} \times
  {_i\mathcal{A}_j^\#}$ cannot be reached by a limit transition.
\end{lem}
\begin{proof}
  Note first that an entry state of a loop of the
  automaton~${_i\mathcal{A}_n}$ cannot be reached by a limit
  transition. This is because Lemma~\ref{lem:loop} just above ensures that
  such a state is reached by a successor transition labeled by the
  letter~$a$ and a state of~${_i\mathcal{A}_n}$ cannot be reached by both
  types of transitions. The same holds for the automata
  ${_i\mathcal{A}_j}$, but not always for the automata
  ${_i\mathcal{A}_j^\#}$. In this latter machine, the added limit
  transition is the only one that may violate the property. This
  transition reaches the state~$j$ which could be the entry state of a
  loop. But, the successor transition leaving $j$ reaches $i+1$ so that,
  for $j$ to be the entry point of a loop, it is necessary that $j= i+1$.
  Then the automaton is said to be \emph{special}. It accepts the set
  $a^\#$ for a letter $a$ and it is pictured in
  Figure~\ref{fig:autom-asharp}. In all other cases, ${_i\mathcal{A}_j^\#}$
  satisfies that an entry state of a loop cannot be reached by a limit
  transition.

  Consider then the Cartesian product ${_i\mathcal{A}_n} \times
  {_i\mathcal{A}_j^\#}$. If a state $(k,k')$ is reached by a limit
  transition, then $k$ and $k'$ are reached by limit transitions in
  ${_i\mathcal{A}_n}$ and ${_i\mathcal{A}_j^\#}$. If
  ${_i\mathcal{A}_j^\#}$ is not special, in each automaton such states are
  not reached by any successor transition. Hence, neither $k$ nor $k'$ are
  entry states of loops. On the other hand, the entry state $(p,p')$ of a
  loop in ${_i\mathcal{A}_n} \times {_i\mathcal{A}_j^\#}$ satisfies either
  $p$ or $p'$ is an entry state of a loop in the corresponding
  automaton. So, the result is proved if ${_i\mathcal{A}_j^\#}$ is not
  special. If it is special, then the Cartesian product is merely a copy of
  ${_i\mathcal{A}_n}$ and the result follows immediately.
\end{proof}

We now introduce the notion of a \emph{trace}. Given a run~$\rho$ of the
automaton ${_i\mathcal{A}_n} \times {_i\mathcal{A}_j^\#}$, the trace~$\tau$ of
the run is defined as follows: each state used in the run occurs in the
trace when it occurs in the run for the first time. It then follows that
such a trace is always finite because it does not contain twice the same
state.  The strong determinism of the automaton ${_i\mathcal{A}_n} \times
{_i\mathcal{A}_j^\#}$ implies that if a state $(k,k')$ occurs in a loop,
then the entry state of the loop occurs in the trace of the run before
$(k,k')$. Similarly, if the state $(q,q')$ is reached from $(k,k')$ by a
forward successor transition, then the state $(q,q')$ occurs just after
$(k,k')$ in the trace of the run. These two remarks directly follow from
the corresponding ones made just above about occurrences of states in the
run of ${_i\mathcal{A}_n} \times {_i\mathcal{A}_j^\#}$.

 We first prove a technical lemma.

\begin{lem}\label{lem:norecur}
  Given a word $x$ and a letter $a$. Let $\rho$ be the run of the automaton
  ${_i\mathcal{A}_n} \times {_i\mathcal{A}_j^\#}$ on $x$ and $\tau$ be the
  trace of this run. Assume that both $\rho$ and $\tau$ have the same last
  state $(k,k')$. Let $(q,q') = (k \cdot a, k' \cdot a)$ be the state of
  ${_i\mathcal{A}_n} \times {_i\mathcal{A}_j^\#}$ reached when reading the
  letter $a$. Then either the state $(q,q')$ is not in $\tau$ or it is the
  entry state of a loop.  In this latter case, the state reached by the
  limit transition associated to the loop is not in $\tau$.
\end{lem}

\begin{proof}
  Assume $(q,q')$ is indeed in the trace $\tau$. Then the run~$\rho$ loops
  over $(q,q')$. Moreover, if $(q,q')$ is not the entry state of the loop,
  this entry state $(p,p')$ occurs in~$\tau$ before $(q,q')$ and all
  states of the loop occur in~$\tau$ after $(p,p')$. In particular, in
  $\tau$, the state that occurs just before $(q,q')$ has to be
  $(k,k')$. This is impossible as soon as the state $(k,k')$ should then
  occur in $\tau$ twice: once just before $(q,q')$ and then at the end of
  $\tau$. This proves that $(q,q')$ is the entry state of the loop. This loop
  gives rise to a limit transition reaching state $(l,l')$. If this state
  occurs in the trace $\tau$, the same argument than the one used for $(q,q')$
  shows that $(l,l')$ has to be the entry state of a loop.  This is
  impossible by Lemma~\ref{lem:entryproduct}.
\end{proof}

We end this paragraph by a technical lemma on the trace of a run.  This
lemma is used later to show that the algorithm indeed terminates.
\begin{lem}\label{lem:jforward}
  Assume that state $i$ is not in a loop and assume that a state $(q,q')$
  occurs in the trace $\tau$ of the run of
  ${_i\mathcal{A}_n} \times {_i\mathcal{A}_j^\#}$.  Then for each
  $i \le p \le q$ there exists at least one state $p'$
  of~${_i\mathcal{A}_j^\#}$ such that $(p,p')$ occurs in the trace~$\tau$.
\end{lem}
\begin{proof}
  If a state~$q$ occurs in the run of~$\mathcal{A}$, so does all states $p
  < q$.  This comes from the way the automaton~$\mathcal{A}$ has been
  built.  On the other hand, if the state~$i$ does not occur in a loop, all
  states accessible from~$i$ are greater than~$i$.  Hence, if $q$ in a run
  of~${_i\mathcal{A}_n}$, all states $p$ such that $i \le p \le q$ also occurs
  in the run. The projection on the first component of the run
  ${_i\mathcal{A}_n} \times {_i\mathcal{A}_j^\#}$ is the run
  of~${_i\mathcal{A}_n}$.  Hence, for each $i \le p \le q$, there exists a
  state~$p'$ of~${_i\mathcal{A}_j^\#}$ such that $(p,p')$ occurs in the run
  ${_i\mathcal{A}_n} \times {_i\mathcal{A}_j^\#}$ before $(q,q')$.  As the same
  is true for the trace of the run, the lemma is proved.
\end{proof}

\subsection{Duplication transformation}\label{sec:duplication}

We define here a transformation~$\tau$ on regular expressions.  Given a
regular expression~$e$, $\tau(e)$ is another regular expression which defines
the same word.  This new expression permits the description of the prime
factorization.  The transformation~$\tau$ is defined by induction on the
expression as follows.

\begin{align*}
  \tau(a) & = a \\
  \tau(ee') & = \tau(e)\tau(e') \\
  \tau(e^\omega) & = \tau(e){\tau(e)}^\omega
\end{align*}

We give below the result of the duplication transformation on
the rational expressions ${(ab)}^\omega$ and ${(a^\omega b)}^\omega a^\omega$.
\begin{exa}
  \begin{align*}
    \tau({(ab)}^\omega) & = \tau(ab){(\tau(ab))}^\omega  = ab{(ab)}^\omega \\
    \tau({(a^\omega b)}^\omega a^\omega) & = \tau({(a^\omega b)}^\omega)\tau(a^\omega) \\
                  & = \tau(a^\omega b)\tau{(a^\omega b)}^\omega\tau(a){\tau(a)}^\omega \\
                  & = \tau(a^\omega)\tau(b){(\tau(a^\omega)\tau(b))}^\omega aa^\omega \\
                  & = aa^\omega b{(aa^\omega b)}^\omega aa^\omega
  \end{align*}
\end{exa}

We let $|e|$ denote the \emph{size} of a regular expression.  This size is
actually the number of letters in $A \cup \{\omega\}$ used in the expression.  We
let also $\depth(e)$ denote its \emph{depth}, that is, the maximum number of
nested $\omega$ in~$e$.  More formally, the size and the depth are inductively
defined as follows.
\begin{alignat*}{2}
  |a|   & = 1          & \qquad \depth(a)   & = 0 \\
  |ee'| & = |e| + |e'| & \qquad \depth(ee') & = \max(\depth(e),\depth(e')) \\
  |e^\omega| & = 1 + |e|    & \qquad \depth(e^\omega) & = 1 + \depth(e)
\end{alignat*}
Note that if $n$ is the size of an expression~$e$, then $n+1$ is the number
of states of the automaton~$\mathcal{A}_e$ constructed in
Section~\ref{sec:singleaut}.
\begin{prop}
  For any regular expression~$e$, the relation
  $|\tau(e)| \le 2^{\depth(e)} |e|$ holds.
\end{prop}
\begin{proof}
  The proof is carried out by induction on the regular expression.  If
  $e = a$, then $\tau(e) = a$, $|e| = |\tau(e)| = 1$ and $\depth(e) = 0$
  and the result holds.  If $e = e'e''$, then $\tau(e) = \tau(e')\tau(e'')$
  and $\depth(e) = \max(\depth(e'),\depth(e''))$.  It follows from the
  induction hypothesis that
  $|\tau(e')| \le 2^{\depth(e')}|e'| \le 2^{\depth(e)}|e'|$ and similarly
  $|\tau(e'')| \le 2^{\depth(e)}|e''|$.  Therefore
  $|\tau(e)| = |\tau(e')| + |\tau(e'')| \le 2^{\depth(e)}(|e'| + |e''|) =
  2^{\depth(e)}|e|$.  If $e = e^{\prime \omega}$, then
  $|\tau(e)| = \tau(e') {\tau(e')}^\omega$ and thus
  $|\tau(e)| = 1 + 2|\tau(e')| \le 1 + 2^{1+\depth(e')}|e'| =
  2^{\depth(e)}|e|$.
\end{proof}

Note that the bound given by the previous proposition is almost sharp.
Consider the expressions ${(e_n)}_{n<\omega}$ defined by induction on~$n$ by $e_0
= a$ and $e_{n+1} = e_n^\omega$.  It can be easily shown by induction on~$n$ that
$|e_n| = n+1$, $\depth(e_n) = n$ and $|\tau(e_n)| = 2^n-1$.

\section{Algorithm}\label{sec:algo}

In this section, we present an algorithm that computes the prime
factorization of a rational word~$x$.  Such a word is given by a rational
expression~$e$.  It turns out that the duplicated expression $\tau(e)$ can
be used to describe the prime factorization of~$x$ by marking main and
secondary cuts of~$x$ in~$\tau(e)$.  Let us illustrate this on the
following example.  Consider the word~$x$ given by the expression
$e = {(bba)}^\omega$.  Then the duplicated expression~$\tau(e)$ is
$bba{(bba)}^\omega$.  The prime factorization of~$x$ is $b^2{(abb)}^\omega$,
that is, $x = u_1^2u_2^\omega$ where the two prime factors are $u_1 = b$
and $u_2 = abb$. It can be given by inserting a marker~$|\!\!|$
(resp.,~$|$) at main (resp., secondary) cuts in the expression $\tau(e)$ as
$|\!\!|b|b|\!\!|a{(bb|a)}^\omega|\!\!|$.  Note that such a marking cannot be
done in the expression~$e$.

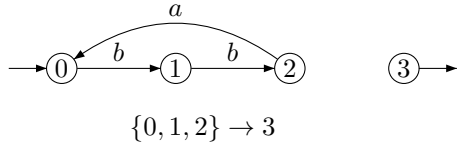
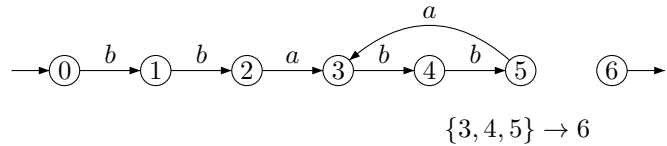
\begin{figure}[htbp]
  \begin{center}
    \gasset{Nw=4,Nh=4,loopdiam=6}
    \begin{picture}(30,17)(0,-8)
    \node[Nmarks=i](0)(0,0){$0$}
    \node(1)(15,0){$1$}
    \node(2)(30,0){$2$}
    \node[Nmarks=f](3)(45,0){$3$}
    \drawedge(0,1){$b$}
    \drawedge(1,2){$b$}
    \drawedge[curvedepth=-6,ELside=r](2,0){$a$}
    \put(9,-8){\makebox(0,0)[l]{$\{0,1,2\} \rightarrow 3$}}
    \end{picture}
    \caption{Automaton for ${(bba)}^\omega$}%
    \label{fig:autom-bbao}
  \end{center}
\end{figure}

The algorithm given below works actually on the
automaton~$\mathcal{A}_{\tau(e)}$ associated to the expression~$\tau(e)$.
Rather than inserting markers in the expression, it distinguishes two
subsets $Q_M$ and~$Q_S$ of states of the automaton~$\mathcal{A}_{\tau(e)}$.
These subsets $Q_M$ and~$Q_S$ correspond to main and secondary cuts
respectively.  As the states of~$\mathcal{A}_{\tau(e)}$ are in one-to-one
correspondence with the positions in~$\tau(e)$, distinguishing states is
the same as inserting markers.  Consider again the expression
$e = {(bba)}^\omega$.  The automaton~$\mathcal{A}_e$ is pictured in
Figure~\ref{fig:autom-bbao} where as the automaton $\mathcal{A}_{\tau(e)}$
is pictured in Figure~\ref{fig:autom-bbaoe}.

\begin{figure}[htbp]
  \begin{center}
    \gasset{Nw=4,Nh=4,loopdiam=6}
    \begin{picture}(75,25)(0,-18)
    \node[Nmarks=i](0)(0,0){$0$}
    \node(1)(12,0){$1$}
    \node(2)(24,0){$2$}
    \node(3)(36,0){$3$}
    \node(4)(48,0){$4$}
    \node(5)(60,0){$5$}
    \node[Nmarks=f](6)(72,0){$6$}
    \drawedge(0,1){$b$}
    \drawedge(1,2){$b$}
    \drawedge(2,3){$a$}
    \drawedge(3,4){$b$}
    \drawedge(4,5){$b$}
    \drawedge[curvedepth=-6,ELside=r](5,3){$a$}
    \put(50,-8){\makebox(0,0)[l]{$\{3,4,5\} \rightarrow 6$}}
    \put(05,-18){\makebox(0,0)[l]{$Q_M = \{0,2,6\}$}}
    \put(35,-18){\makebox(0,0)[l]{$Q_S = \{1,5\}$}}
    \end{picture}
    \caption{Automaton for $\tau({(bba)}^\omega) = bba{(bba)}^\omega$}%
    \label{fig:autom-bbaoe}
  \end{center}
\end{figure}

The algorithm works on the second automaton.  The subsets $Q_M$ and~$Q_S$
are respectively $Q_M = \{0,2,6\}$ and $Q_S = \{1,5\}$. States $0$ and~$6$
are visited at the two trivial cuts $\varepsilon\cdot {(bba)}^\omega$ and
${(bba)}^\omega \cdot \varepsilon$.  State~$2$ is visited at the main cut
$b^2 \cdot {(abb)}^\omega$.  The unique visit of state~$1$ occurs at the
secondary cut $b \cdot b{(abb)}^\omega$.  Visits of state~$5$ occur at
secondary cuts of the form $b^2{(abb)}^n \cdot {(abb)}^\omega$ for an
integer~$n$.  States $3$ and~$4$ are always visited at cuts that are
neither main nor secondary.  It should be noted that such a separation
between states cannot be done on the first automaton.  Indeed, in the first
automaton, the first visit of state~$1$ occurs at the secondary cut
$b \cdot b{(abb)}^\omega$.  All the other visits occur at cuts of the form
${(bba)}^n b \cdot {(bab)}^\omega$ which are not secondary.  This is why the
duplication operator~$\tau$ has been introduced.  It can be easily seen
that distinguishing states of~$\mathcal{A}_{\tau(e)}$ is indeed the same as
inserting markers in the expression~$\tau(e)$.  The main result is that
what happens on the example is general: some states of the automaton
$\mathcal{A}_{\tau(e)}$ correspond to main cuts and some of them correspond
to secondary cuts.  The algorithm computes the two subsets $Q_M$ and~$Q_S$.
We can now state the main result of this section.

\begin{thm}
  Given a rational word~$x$ denoted by a regular expression~$e$, there are
  two subsets $Q_M$ and~$Q_S$ of states of~$\mathcal{A}_{\tau(e)}$ such that
  the main and secondary cuts are exactly those mapped to states in $Q_M$
  and~$Q_S$ by the run labeled by~$x$.  Furthermore, these subsets can be
  computed in polynomial time in the size of~$\tau(e)$.
\end{thm}

\subsection{Description}\label{sec:algodesc}

The algorithm is essentially inspired by its counterpart used in the
classical case of finite words~\cite{Duval80}. In this case, three
variables $i$, $j$ and~$k$ representing positions in the word are used. The
variable $i$ contains a position such that the prefix of the finite word
ending at this position is already factorized. The variable $j$ contains a
position greater than~$i$ such that the factor between positions given by
$i$ and~$j$ is the possible next prime factor. The variable $k$ is greater
than~$j$ and contains the current position in the finite word. Moreover, to
make the classical algorithm easier to understand, a fourth variable~$k'$
can be introduced. It contains a position in the possible next prime
factor, this position ranges between the positions $i$ and~$j$. The
classical algorithm is then directed by the comparison of the letters just
after the positions $k$ and~$k'$.

Given a rational word $x$ represented by an expression~$e$, the algorithm
presented here works on the automaton $\mathcal{A}_{\tau(e)}$ denoted
$\mathcal{A}$. It also uses four variables $i$, $j$, $k$ and~$k'$.
Theses variables do not contain positions in the word~$x$, but rather
states of $\mathcal{A}$ for the first three ones and a state of
${_i\mathcal{A}_j^\#}$ for the last one~$k'$. The algorithm produces two
subsets of states of $\mathcal{A}$, the first one $Q_M$ contains states
corresponding to main cuts and the second one $Q_S$ contains states
corresponding to secondary cuts. It also uses a list of states of the
Cartesian product called the history.  This history is written below using
angle brackets.  It is the trace of the run of the product automaton on the
the factor ${_{i}x_k}$. The algorithm is directed by the state of
${_i\mathcal{A}_n} \times {_i\mathcal{A}_j^\#}$ given by the last state of the
history. This pair of states is called the leading pair. It is formed of
the states contained in the variables $k$ and~$k'$.  As in the finite case,
the algorithm is directed by the comparison of the letters labeling the
unique successor transition leaving $k$ in~${_i\mathcal{A}_n}$ and~$k'$
in~${_i\mathcal{A}_j^\#}$. This leads to the three cases described below
according to the fact that the two letters are the same (case 1), or the
first is greater than the second (case 2) or the first is smaller than the
second (case 3). In case 2, the content of variable~$j$ is changed and
a new automaton ${_i\mathcal{A}_j^\#}$ is considered. In case 3, the
next prime factor is found and variables $i$, $j$, $k$ and~$k'$ are
reinitialized to new values and a new automaton ${_i\mathcal{A}_j^\#}$ is
considered.  Moreover, both sets $Q_M$ and~$Q_S$ are updated.

The algorithm starts with $i=0$, $j=1$, $k=1$ and $k'=0$. The history is
just the list $\langle(k,k')\rangle$, so that the leading pair is
$(k,k') = (1,0)$. The sets $Q_M$ and~$Q_S$ are initialized to $Q_M = \{0\}$
and $Q_S = \emptyset$. Now assume that variables $i$, $j$, $k$ and~$k'$ are
known and that the history, ending with $(k,k')$, is known as well. Assume
too that the sets $Q_S$ and~$Q_M$ are known. Then, as announced just above,
the behavior of the algorithm falls in one of the three cases given below.

Both automata ${_i\mathcal{A}_n}$ and ${_i\mathcal{A}_j^\#}$ are strongly
deterministic.  From any state, there is unique successor transition
leaving it.  In the description below, its label is called the letter
\emph{leaving the state}.  Moreover, it is assumed that a fake end-maker,
which is smaller than any other letter, is the letter leaving the accepting
state~$n$ of~${_0\mathcal{A}_n}$.

\begin{itemize}[align=left]
\item[Case 1:] The letters leaving $k$ and $k'$ are the same letter~$a$.
  Compute the new pair $(k\cdot a,k'\cdot a)$

  The case now splits in three sub-cases.

  \begin{itemize}[align=left]
  \item[Case 1a:] The new pair is not in the history.  Then it is just
    added to the history and the algorithm goes on.
  \item[Case 1b:] The new pair is in the history and the loop
    in~${_i\mathcal{A}_j^\#}$ from~$k'\cdot a$ to $k'\cdot a$ does not
    visit state~$j$.  The trace correspond to two loops $P$ and~$P'$ in
    ${_i\mathcal{A}_n}$ and ${_i\mathcal{A}_j^\#}$.  The pair
    $(1+\max P,1+\max P')$ is then added to the history and the algorithm
    goes on.
  \item[Case 1c:] The new pair is in the history and the loop
    in~${_i\mathcal{A}_j^\#}$ from~$k'\cdot a$ to $k'\cdot a$ visits
    state~$j$.  The trace correspond to two loops $P$ and~$P'$ in
    ${_i\mathcal{A}_n}$ and ${_i\mathcal{A}_j^\#}$ where $P'$ contains $j$.
    The pair $(1+\max P,j)$ is then added to the history and the algorithm
    goes on.
  \end{itemize}
  Note that Case 1c is similar to case 1b with the convention that $j =
  j+1$ in ${_i\mathcal{A}_j^\#}$.

\item[Case 2:] The letter~$b$ leaving $k$ is greater than the letter~$a$
  leaving $k'$.  The possible next prime factor has to be changed.

  There are two cases.
  \begin{itemize}[align=left]
  \item[Case 2a:] If $k\cdot b$ does not occur in the history, $j$ is set to
    $k\cdot b$ and the leading pair is $(i,j)$. The variable~$k$ is set to $j$
    and $k'$ is set to~$i$.  The history is erased and reset to the list
    reduced to $(k,k')$.
  \item[Case 2b:] If $k\cdot b$ occurs in the history, let $m$ be the largest
    state in the history.  Then the variable~$j$ is set to $m+1$. The
    variable~$k$ is set to $j$ and $k'$ is set to~$i$.  The history is
    erased and reset to the list reduced to $(k,k')$.
  \end{itemize}
  The automaton ${_i\mathcal{A}_j^\#}$ is reconstructed and the algorithm
  goes on.

  Note that in both cases, the new value of~$j$ is greater than the
  previous one, due to Lemma~\ref{lem:jforward}.

\item[Case 3:] The letter~$a$ leaving $k$ is smaller than the letter~$b$
  leaving $k'$.  This case includes the case where $k$ is the accepting
  state~$n$ and $a$ the fake end-marker.

  State $j$ and all states~$q$ such that $(q,j)$ occurs in the history are
  added to~$Q_S$.  The greatest added state is removed from~$Q_S$ and added
  to~$Q_M$ and variable~$i$ is set to this state.

  Let $c$ be the label of the unique successor transition from this
  new~$i$.  The indices $j$ and~$k$ are set to~$i\cdot c$.  The
  variable~$k'$ is set to~$i$.  The leading pair is then
  $(k = i\cdot c, k' = i)$.  The history is reset to the list reduced to
  the pair $(k,k')$.  The automaton ${_i\mathcal{A}_j^\#}$ is reconstructed
  and the algorithm goes on.
\end{itemize}
The algorithm stops when $i$ is the accepting state~$n$
of~${_0\mathcal{A}_n}$.  A formal description of the algorithm is given
below, after the following example.

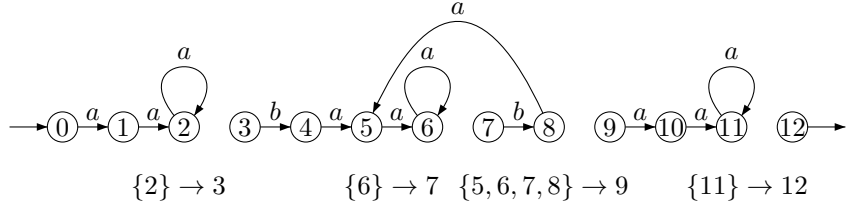
\begin{figure}[htbp]
  \begin{center}
    \gasset{Nw=4,Nh=4,loopdiam=6}
    \begin{picture}(90,25)(0,-8)
    \node[Nmarks=i](0)(0,0){$0$}
    \node(1)(8,0){$1$}
    \node(2)(16,0){$2$}
    \node(3)(24,0){$3$}
    \node(4)(32,0){$4$}
    \node(5)(40,0){$5$}
    \node(6)(48,0){$6$}
    \node(7)(56,0){$7$}
    \node(8)(64,0){$8$}
    \node(9)(72,0){$9$}
    \node(10)(80,0){$10$}
    \node(11)(88,0){$11$}
    \node[Nmarks=f](12)(96,0){$12$}
    \drawedge(0,1){$a$}
    \drawedge(1,2){$a$}
    \drawloop(2){$a$}
    \drawedge(3,4){$b$}
    \drawedge(4,5){$a$}
    \drawedge(5,6){$a$}
    \drawloop(6){$a$}
    \drawedge(7,8){$b$}
    \drawedge[ELside=r,curvedepth=-14](8,5){$a$}
    \drawedge(9,10){$a$}
    \drawedge(10,11){$a$}
    \drawloop(11){$a$}
    \put(9,-8){\makebox(0,0)[l]{$\{2\} \rightarrow 3$}}
    \put(37,-8){\makebox(0,0)[l]{$\{6\} \rightarrow 7$}}
    \put(52,-8){\makebox(0,0)[l]{$\{5,6,7,8\} \rightarrow 9$}}
    \put(82,-8){\makebox(0,0)[l]{$\{11\} \rightarrow 12$}}
    \end{picture}
    \caption{Automaton for $aa^\omega b{(aa^\omega b)}^\omega aa^\omega$}%
    \label{fig:autom-aaobaaoboaao}
  \end{center}
\end{figure}

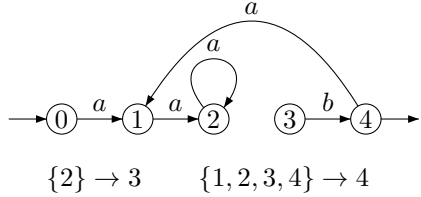
\begin{figure}[htbp]
  \begin{center}
    \gasset{Nw=4,Nh=4,loopdiam=6}
    \begin{picture}(30,25)(0,-8)
    \node[Nmarks=i](0)(0,0){$0$}
    \node(1)(10,0){$1$}
    \node(2)(20,0){$2$}
    \node(3)(30,0){$3$}
    \node[Nmarks=f](4)(40,0){$4$}
    \drawedge(0,1){$a$}
    \drawedge(1,2){$a$}
    \drawloop(2){$a$}
    \drawedge(3,4){$b$}
    \drawedge[ELside=r,curvedepth=-13](4,1){$a$}
    \put(-2,-8){\makebox(0,0)[l]{$\{2\} \rightarrow 3$}}
    \put(18,-8){\makebox(0,0)[l]{$\{1,2,3,4\} \rightarrow 4$}}
    \end{picture}
    \caption{Automaton ${_0\mathcal{A}_4^\#}$}%
    \label{fig:autom-0a4}
  \end{center}
\end{figure}

We now give the execution of the algorithm on the expression
$e = {(a^\omega b)}^\omega a^\omega$.  As seen before, its duplication
transformation $\tau(e)$ is equal to
$aa^\omega b{(aa^\omega b)}^\omega aa^\omega$.  The corresponding automaton
has $13$ states and is pictured in Figure~\ref{fig:autom-aaobaaoboaao}. In
the sequel, all descriptions of the automata~${_i\mathcal{A}_j^\#}$ but
${_0\mathcal{A}_4^\#}$ are skipped.  The automaton ${_0\mathcal{A}_4^\#}$
is pictured in Figure~\ref{fig:autom-0a4}.

The algorithm starts with $i = 0$, $j = k = 1$ and $k' = 0$.  The history
is just the list $\langle(1,0)\rangle$ made of this single pair and thus
the leading pair is thus $(1,0)$.  The sets $Q_M$ and~$Q_S$ are
$Q_M = \{0\}$ and $Q_S = \emptyset$.

The letter leaving states $k = 1$ and $k' = 0$ is the letter~$a$. Hence,
this is case~1. The pair $(k\cdot a, k'\cdot a)$ is the pair $(2,1)$, since
$k'$ is a state of~${_0\mathcal{A}_1^\#}$ pictured in
Figure~\ref{fig:autom-asharp}.  Since this pair does not occur in the
history, this is case~1a and the history becomes
$\langle(1,0),(2,1)\rangle$.  The new leading pair is the pair $(2,1)$.

The letter leaving states $k = 2$ and $k' = 1$ is the letter~$a$.  Hence,
this is again case~1.  The pair $(k\cdot a, k'\cdot a)$ is the pair $(2,1)$
that already occurs in the history, detecting a loop visiting state
$j = 1$.  Hence, this is case~1c.  The pair added to the history is
$(3,1)$.  The history becomes $\langle(1,0),(2,1),(3,1)\rangle$.

The letter leaving states $k = 3$ and $k' = 1$ are respectively the letters
$b$ and~$a$.  This is case~2.  As $4 = k\cdot b$ does not occur in the
history, this is case~2a.  Then $j$ is set to $k\cdot b = 4$.  Variable~$k$
is set to~$4$ and $k'$ is reset to~$0$.  The history is reset to
$\langle(4,0)\rangle$.  The algorithm uses the
automaton~${_0\mathcal{A}_4^\#}$ pictured in Figure~\ref{fig:autom-0a4}.

The letter leaving states $k = 4$ and $k' = 0$ is the letter~$a$ giving
rise to the pair $(5,1)$.  This is case~$1a$ and this history becomes
$\langle(4,0),(5,1)\rangle$.

The letter leaving states $k = 5$ and $k' = 1$ is the letter~$a$ giving
rise to the pair $(6,2)$.  This is case~$1a$ and this history becomes
$\langle(4,0),(5,1),(6,2)\rangle$.

The letter leaving states $k = 6$ and $k' = 2$ is the letter~$a$ giving
rise to the pair $(6,2)$ which is already in the history.  State $j = 4$ is
not visited in the loop.  This is case~1b.  The new pair is then $(7,3)$
and this history becomes $\langle(4,0),(5,1),(6,2),(7,3)\rangle$.

The letter leaving states $7$ and~$3$ is the same letter~$b$ giving rise
to the pair $(8,4)$ which is added to the history.  This history is then
$\langle(4,0),(5,1),(6,2),(7,3),(8,4)\rangle$.

The letter leaving~$k = 8$ is the letter~$a$.  The letter leaving $k' = 4$
in the automaton ${_0\mathcal{A}_4^\#}$ is also the letter~$a$: state~$4$
simulates state~$0$ (see Figure~\ref{fig:autom-0a4}).  This gives rise to
the new pair $(5,1)$.  There is a loop since $(5,1)$ already occurs in the
history.  State $j = 4$ is indeed visited by the pair $(8,4)$ in the
history.  Thus, this is case~1c.  Hence the new pair is $(9,4)$.  This
history becomes $\langle(4,0),(5,1),(6,2),(7,3),(8,4),(9,4)\rangle$.

The letter leaving states $9$ and~$4$ is the same letter~$a$ giving rise
to the pair $(10,1)$ which is added to the history.  This history is then
\begin{displaymath}
  \langle(4,0),(5,1),(6,2),(7,3),(8,4),(9,4),(10,1)\rangle.
\end{displaymath}
The letter leaving states $10$ and~$1$ is the same letter~$a$ giving rise
to the pair $(11,2)$ which is added to the history.  This history is then
\begin{displaymath}
  \langle(4,0),(5,1),(6,2),(7,3),(8,4),(9,4),(10,1),(11,2)\rangle.
\end{displaymath}
The letter leaving states $k = 11$ and $k' = 2$ is the letter~$a$ giving
rise to the pair $(11,2)$ which is already in the history.  State $j = 4$
is not visited in the loop.  This is case~1b.  The new pair is then
$(12,3)$ and this history becomes
\begin{displaymath}
  \langle(4,0),(5,1),(6,2),(7,3),(8,4),(9,4),(10,1),(11,2),(12,3)\rangle.
\end{displaymath}
Due to the convention, the letter leaving state~$12$ is the fake right-end
marker which is smaller than all other letters.  This is thus Case~3.
States $4$, $8$ and~$9$ are added to~$Q_S$ and $9$ is removed from~$Q_S$
and added to~$Q_M$.  So $Q_M = \{0,9\}$ and $Q_S = \{4,8\}$.  Variables $i$
and~$k$ are set to $9$ and variables $j$ and~$k'$ are set to~$10$.  The
history is reset to $\langle(10,9)\rangle$.  The algorithm uses the
automaton~${_9\mathcal{A}_{10}^\#}$ which is similar to the
automaton~${_0\mathcal{A}_1^\#}$ pictured in Figure~\ref{fig:autom-asharp}.

The letter leaving states $10$ and~$9$ is the same letter~$a$ giving rise
to the pair $(11,10)$ which is added to the history.  This history is then
$\langle(10,9),(11,10)\rangle$.

The letter leaving states $k = 11$ and $k' = 10$ is the letter~$a$ giving
rise to the same pair $(11,10)$.  As $j=10$ is visited in the loop,
this is case 1c.  The pair $(12,10)$ is added to the history which becomes
$\langle(10,9),(11,10),(12,10)\rangle$.

Due to the convention, the letter leaving state~$12$ is the fake right-end
marker which is smaller than all other letters.  This is thus Case~3.
State $j = 10$ and states $11$ and~$12$ are added to~$Q_S$.  State~$12$ is
removed from~$Q_S$ and added to~$Q_M$.  So $Q_M = \{0,9,12\}$ and
$Q_S = \{4,8,10,11\}$.  Finally variable~$i$ is set to $12$ and the
algorithm stops. This gives the factorization of the word in
$u_1^\omega u_2^\omega$ where the primes words $u_1$ and~$u_2$ are
$u_1 = a^\omega b$ and $u_2 = a$.

This example shows that several points have to be proved.  We give here
three such points.  Each time a state is added to $Q_M$ or~$Q_S$, all
visits of that state occur at main or secondary cuts.  As soon as a
state~$q$ is added to~$Q_M$, all states visited after~$q$ are greater than
$q$.  The algorithm terminates and computes the prime factorization of the
word.

\renewcommand{\algorithmicreturn}{\textbf{Output}}
\begin{algorithm}
\caption{\textsc{LyndonFactorize}}
\algsetup{indent=2em}
\begin{algorithmic}[1]
  \STATE Input: automaton $(\{0,\ldots,n\},A,E,\{0\},\{n\})$.
  \STATE $i \leftarrow 0$, $j \leftarrow 1$, $k \leftarrow 1$, $k' \leftarrow 0$, $H = (i,j) = (0,1)$, $Q_M \leftarrow
  \{0\}$, $Q_S \leftarrow \emptyset$
  \WHILE{$k < n$}
    \IF{$a_k = a_{k'}$}
      \STATE $k \leftarrow k \cdot a_k$ in $\mathcal{A}$ and $k' \leftarrow k' \cdot a_{k'}$ in ${_i\mathcal{A}_j^\#}$
      \IF{$(k,k')$ occurs in~$H$}
        \IF{$j$ does not occur since the previous visit of $(k,k')$}
          \STATE $k \leftarrow  \max \{q \mid\exists q' \;\; (q,q') \text{ occurs in $H$ after $(k,k')$}\}$
          \STATE $k' \leftarrow \max \{q' \mid\exists q \;\; (q,q') \text{ occurs in $H$ after $(k,k')$}\}$
        \ELSE
          \STATE $k \leftarrow \max \{q \mid\exists q' \;\; (q,q') \text{ occurs in $H$ after $(k,k')$}\}$
          \STATE $k' \leftarrow j$
        \ENDIF
      \ENDIF
      \STATE $H \leftarrow H \cdot (k,k')$
    \ELSIF{$a_k \gtalp a_{k'}$}
      \IF{$k\cdot a_k$ does not occur in~$H$}
        \STATE $j \leftarrow k\cdot a_k$
      \ELSE
        \STATE $j \leftarrow \max \{q \mid\exists q' \;\; (q,q') \in H\} + 1$
      \ENDIF
      \STATE $k \leftarrow j$, $k' \leftarrow i$, $H \leftarrow (i,j)$
    \ELSE
      \STATE $Q_S \leftarrow Q_S \cup \{j\} \cup \{ q \mid(q,j) \in H \}$.
      \STATE $i \leftarrow \max Q_S$, $Q_S \leftarrow Q_S \setminus\{i\}$, $Q_M \leftarrow Q_M \cup \{i\}$
      \STATE $j \leftarrow i\cdot a_i$, $k \leftarrow j$, $k' \leftarrow i$, $H \leftarrow (i,j)$
    \ENDIF
  \ENDWHILE
  \RETURN $Q_M$ and $Q_S$
\end{algorithmic}
\end{algorithm}

\subsection{Additional properties of prime words}\label{sec:addprop}

We first state three properties on prime words that are used to prove that
the algorithm given just above is correct.

\begin{lem}
  Let $u$ be a prime word and let $xa$ be a prefix of~$u$ where $x$ is a
  word and $a$ is a letter.  If $b$ is a letter such that $a \ltalp b$,
  then the word~$xb$ is prime and satisfies $u \ltlex xb$.
\end{lem}
\begin{proof}
  Since $xa$ is a prefix of~$u$, there exists a word~$y$ such that $u =
  xay$.  We first show that any suffix of~$xb$ is greater than~$xb$.  A
  suffix of $vb$ is either the letter~$b$ or has the form $x'b$ where $x'$
  is a non-empty suffix of~$x$.  In the former case, since $a$ occurs
  in~$u$, the first letter~$a'$ of~$u$ must satisfy $a' \ltalp a$.
  Otherwise the suffix $ay$ would satisfy $ay \ltstr u$ and this would
  contradict the fact that $u$ is prime.  It follows then that $u \ltstr
  b$.  In the latter case, note that $|x'| \le |x|$ since $x'$ is a suffix
  of~$x$. The case $x' = x$ is trivial and we assume therefore that $x' \neq
  x$.  Since $u$ is prime, the suffix $x'ay$ of~$u$ satisfies $u \lelex
  x'ay$, that is, either $u \ltstr x'ay$ or $u = x'ay$.  This implies that
  either $x'a \ltpre u$ or $u \ltstr x'a$.  In both cases, one has $u
  \ltstr x'b$ and thus $yb \ltstr x'b$ since $|x'| \le |x|$.

  It remains to show that $xb$ is primitive.  If $xb$ is not primitive, by
  Lemma~\ref{lem:power}, it is equal to $z^\alpha$ for some word~$z$ and
  some ordinal~$\alpha$ which is a power of~$\omega$.  This is not possible
  since $xb$ has a last letter.
\end{proof}

The following corollary is directly obtained by combining the previous
lemma with Proposition~\ref{pro:ualphavprime}.
\begin{cor}\label{cor:ualphaxbprime}
  Given a prime word $u = xay$ and a letter $b$ such $a \ltalp b$, the word
  $u^\alpha xb$ is prime for any ordinal~$\alpha$.
\end{cor}

The next lemma states that given a prime word~$u$ and a word~$x$ such that
$x \ltstr u$, the prime factorization of $u^\alpha x$ is made of $\alpha$
copies of~$u$ followed by the prime factorization of~$x$.

\begin{lem}\label{lem:nobacktrack}
  Let $u$ be a prime word and let $x$ a be a word such that $x \ltstr u$.
  For any ordinal~$\alpha$, the prime factorization of the word
  $u^\alpha x$ has the form $u^\alpha \xi$ where $\xi$ is the prime
  factorization of~$x$.
\end{lem}
\begin{proof}
  Since the prime factorization is unique by Theorem~\ref{thm:main}, it
  suffices to show that the sequence $u^\alpha \xi$ is indeed a densely
  non-increasing sequence of prime words.  It is clear that this sequence
  only contains prime words.  Let $v_0$ the first prime word that occurs in
  the prime factorization~$\xi$ of~$x$.  We claim that this word~$v_0$
  satisfies $v_0 \lelex u$.  Since $x \ltstr u$, there exist words $y$,
  $u'$ and~$x'$ and letters $a$ and~$b$ such that $x = yax'$, $u = ybu'$
  and $a \ltalp b$.  Note that $v_0$ is a prefix of~$x$.  If
  $|v_0| \le |y|$, then $v_0$ is a prefix of~$y$ and thus a prefix of~$u$
  which implies $v_0 \lelex u$.  If $|v_0| > |y|$, then $ya$ is a prefix
  of~$v_0$ and thus $v_0 \ltstr u$ which implies $v_0 \lelex u$.

  Since $v_0 \lelex u$, the sequence $u^\alpha \xi$ is indeed densely
  non-increasing and it is then the prime factorization of $u^\alpha x$.
\end{proof}

\subsection{Invariants}\label{sec:invariants}

In order to prove that the algorithm is correct, we prove that the
following six invariants always hold during its execution.  The main
invariant is the first one that guarantees the correctness of the
algorithm.  The five others are more technical invariants used to
prove the first one.

\begin{enumerate}
\item The prime factorization of ${_0x_i}$ is
  ${_0x_i} = u_1^{\alpha_1} \cdots u_r^{\alpha_r}$.  Its main cuts are
  exactly those mapped to states in~$Q_M$.  Its secondary cuts are exactly
  those mapped to states in~$Q_S$.
\item The state~$i$ is not in a loop.
\item The prime word $u_r$ satisfies $u_r \gtlex {_{i}x_n}$.
\item The word ${_{i}x_j}$ is prime.
\item Let ${_{(i,i)} x_{(k,k')}}$ be the word labelling the run in
  ${_i\mathcal{A}_n} \times {_i\mathcal{A}_j^\#}$ from $(i,i)$ to~$(k,k')$
  without visiting $(k,k')$.  Then ${_{(i,i)} x_{(k,k')}} = {_{i}x_j^\beta}y$
  with $y \ltpre {_{i}x_j}$.
\item The history $H$ contains the trace of a run in
  ${_i\mathcal{A}_n} \times {_i\mathcal{A}_j^\#}$ and thus contains no
  repetition.
\end{enumerate}

We now show that these six invariants do hold.  Notice that Invariants 1,2
and~3 only depend on~$i$ and not on $j$ and~$k$.  Similarly, Invariant~4
only depends on $i$ and~$j$ and not on~$k$.

At the beginning of the algorithm, $i = 0$ and $j = k = 1$.  The invariants
1 and~3 hold by vacuity.  Invariant~2 is true since $i = 0$.  Invariant~4
holds since ${_0x_i} = {_{i}x_j}$ is just the first letter of~$x$.
Invariants 5 and~6 are also trivially true.

Now assume that the variables $i$, $j$, $k$ and~$k'$ and the history~$H$
are known.  The last pair in~$H$ is the leading pair $(k,k')$.

We prove that after each iteration of the main while loop of the algorithm,
the six invariants still hold.  Each iteration falls in one of the three
cases already described.  We now consider each case.

\paragraph{Case 1.}

In this case, neither $i$ nor $j$ are changed.  Hence, Invariants 1,2,3
and~4 remain obviously true.

The two automata ${_i\mathcal{A}_n}$ and ${_i\mathcal{A}_j^\#}$ are in
states $k$ and $k'$ and both successor transitions leaving $k$ and~$k'$ are
labelled the same letter~$a$.  Then this is either case 1a, 1b or~1c.

case 1a: the new pair $(k\cdot a, k'\cdot a)$ is not in the history.  Then it is
added to it. The history remains the trace of the run.  The suffix~$y$
becomes $ya$ which is still a prefix of ${_{i}x_j}$.  If $k'\cdot a = j$, then
$ya = {_{i}x_j}$, the exponent~$\beta$ is increased by~$1$, and $y$ becomes the
empty word.  Hence, Invariants 5 and~6 hold.

case 1b: the new pair $(k\cdot a, k'\cdot a)$ is already in the history~$H$
and, after its first occurrence, state $j$ does not occur as a first
component of a pair. Note that, then, its sure that $k'\cdot a \neq j$. An
internal loop of ${_i\mathcal{A}_j^\#}$ has been reached, that is, a loop
that is already a loop of~${_i\mathcal{A}_j}$ (i.e.\ using none of the two
added transitions). This loop is also a loop of~${_i\mathcal{A}_n}$. The
repetition $\omega$ times of the two loops leads to a use of limit
transitions in the two automata. Let $q$ and $q'$ be the two states reached
by these two limit transitions.  Then the word ${_{i}x_j}$ can be factorized
as ${_{i}x_j} = x_1x_2^\omega x_3$ where $x_2$ is the label of the loop. The
new read prefix is then of the form $yx_2^\omega$. Then Invariant~5 holds.
The only invariant possibly violated is Invariant~6: it remains to be shown
that the new pair~$(q,q')$ added to the history does not already occur in
it.  This is guaranteed by Lemma~\ref{lem:norecur}.

case 1c: the new pair $(k\cdot a, k'\cdot a)$ is already in the history~$H$
and, after its first occurrence, state $j$ does occur as a second component
of a pair. This means that a loop labeled by a power~$\gamma$ of a
conjugate the word ${_{i}x_j}$ has been read in
${_i\mathcal{A}_n} \times {_i\mathcal{A}_j^\#}$.  Then this leads to the
use a limit transition in ${_i\mathcal{A}_n}$ and to the use of the added
limit transition in~${_i\mathcal{A}_j^\#}$. The reached state in
${_i\mathcal{A}_n}$ is $\ell$ where $\ell-1$ is the maximum state in the
loop of~${_i\mathcal{A}_n}$ and the reached state in~${_i\mathcal{A}_j^\#}$
is state~$j$. The new prefix read is then of the form
${_{i}x_j}^{\beta+\lambda\cdot\omega}$ and $y$ becomes empty. As in case 1b,
Invariant~5 holds and the only invariant possibly violated is then
invariant~6.  It remains to be shown that the new pair $(q,j)$ added to the
history do not already occur in it.  Again Lemma~\ref{lem:norecur} gives
the result.

\paragraph{Case 2.}

In this case, $i$ is not changed.  Hence, Invariants 1,2 and~3 remain
obviously true.

The two automata ${_i\mathcal{A}_n}$ and ${_i\mathcal{A}_j^\#}$ are in
states $k$ and $k'$.  Let $b$ (respectively $a$) be the letter labelling
the successor transition leaving~$k$ (respectively $k'$) with $a \ltalp b$.
Let $q$ and $q'$ be the states $k \cdot b$ and $k' \cdot a$.  The word
${_{i}x_j}$ can be factorized ${_{i}x_j} = x_1ax_2$ where $x_1 = y$ and the
word ${_{(i,i)} x_{(k,k')}}$ is followed by letter~$b$.  By
Corollary~\ref{cor:ualphaxbprime}, the word ${(x_1ax_2)}^\beta x_1b$ is
prime.  Let $j'$ be the state $k\cdot b$. Depending on the value of $j'$,
this is either Case~2a or Case~2b.

Case 2a: the state $j'$ never occurred up to now in the run performed by
${_i\mathcal{A}_n}$. Then variable~$j$ is set to $j'$.  Invariant~4 holds
since ${(x_1ax_2)}^\beta x_1b$ is prime. Invariant~5 trivially holds. Since
the history is reset to the single pair
$\langle(i,j)\rangle = \langle(k,k')\rangle$, Invariant~6 holds.

Case 2b: the state $j'$ already occurred in the run performed by
${_i\mathcal{A}_n}$. This means that this automaton has entered a loop. Let
$\ell-1$ be the largest state visited in the loop. The
automaton~${_i\mathcal{A}_n}$ can use a limit transition to reach
state~$\ell$. As in case~2a, the word ${(x_1ax_2)}^\beta x_1b$ is prime.
This word ends with a due to the loop over $j'$ and the duplication.  By
Lemma~\ref{lem:squareend}, the word ${_{i}x_p}$ is prime.  This ensures that
Invariant~4 holds as $j$ is set to~$q$.  Invariant~5 trivially holds. Since
the history is reset to the single pair
$\langle(i,\ell)\rangle = \langle(k,k')\rangle$, Invariant~6 holds.

\paragraph{Case 3.}

The two automata ${_i\mathcal{A}_n}$ and ${_i\mathcal{A}_j^\#}$ are in
states $k$ and~$k'$.  Let $a$ (respectively $b$) be the letter labelling
the successor transition leaving~$k$ (respectively $k'$) with $a \ltalp b$.
Let $q$ and $q'$ be the states $k \cdot a$ and $k' \cdot b$.  During the
proof of the invariants, we let $i$ and $i'$ denote respectively the old
and new values of the variable~$i$.  Let us recall that $i'$ is the largest
state in $\{j \} \cup \{q \mid(q,j) \in H\}$.

We begin with Invariant~3.  The word ${_{i}x_j}$ is the prime factor
$u_{r+1}$.  It can be factorized as $u_{r+1} = x_1bx_2$ where $x_1 = y$ and
the word ${_{(i,i)} x_{(k,k')}}$ is followed by letter~$a$.  This ensures
that $u_{r+1} \gtlex x_1a \gtlex {_{i'}x_n}$.  This proves Invariant~3.

We continue with the first part Invariant~1, namely that the prime
factorized of ${_0x_{i'}}$ is
$u_1^{\alpha_1} \cdots u_r^{\alpha_r}u_{r+1}^{\alpha_{r+1}}$.  The word
$u_{r+1}$ is a prefix of ${_{i}x_n}$.  It satisfies therefore
$u_{r+1} \ltlex {_{i}x_n} \ltlex u_r$ by Invariant~1 applied to the old
value~$i$.  This ensures that the prime factorization of ${_0x_{i'}}$ is
${_0x_{i'}} = u_1^{\alpha_1} \cdots u_r^{\alpha_r}u_{r+1}^{\alpha_{r+1}}$
where $\alpha_{r+1} = \beta$.

We now prove Invariant 2, namely that $i'$ is not in a loop. The proof is
by contradiction. So, assume $i'$ is in a loop of the automaton
${_0\mathcal{A}_n}$. The loop has an entry state $p$ and is labeled by a
word $z'$. By Lemma~\ref{lem:loop}, $z'$ has a last letter $c$ and can be
factorized as $z' =zc$ and $x = x_1c{(zc)}^\omega x_2$. Moreover, due to the
duplication, $x$ can be factorized as $x= x_1czc{(zc)}^\omega x_2$ so that
the loop is entered at the cut $x_1czc \cdot {(zc)}^\omega x_2$. As $i'$ is
in the loop, it gives rise to a cut of the form
$x_1czcy_1 \cdot y_2{(zc)}^\omega y_3$, so that the cut associated to $i'$ do
not occur in the occurrence of $cz$ which follows the prefix $x_1$. But, by
Corollaries~\ref{cor:cutsproduct} and~\ref{cor:cutsomega}, all main cuts
of~$x$ must occur either in the prefix $x_1cz$ or in the suffix $x_2$ and
this is a contradiction.

We come back to Invariant~1.  From Invariant~2, the state~$i'$ occurs only
once in the run and therefore there is only one cut mapped to~$i'$ and it
is indeed a main cut.  In order to prove Invariant~1, it remains to prove
that secondary cuts are exactly those mapped to states in~$Q_S$. As any
secondary cut is mapped to a state in~$Q_S$, it is enough to prove that any
cut mapped to a state in $Q_S$ is secondary.  Consider then a state
$p \in Q_S$. If $p$ is not in a loop, it occurs only once in the run and
the result is immediate. So, from now on, state $p$ is assumed to be in a
loop labelled by $z'$. Then we first prove that the first cut of
${_0x_{i'}}$ mapped to $p$ is a secondary cut of ${_0x_{i'}}$. Consider a
cut ${_0x_{i'}} = x_1 \cdot x_2$ mapped to $p$. Then ${_0x_{i'}}$ can be
factorized as ${_0x_{i'}} = y_1 z^{\prime \omega} y_2$. Then, just as
above, $z'$ has a last letter $c$ such that $z'=zc$ and due to the
duplication and to Lemma~\ref{lem:loop}, the factorization can be written
${_0x_{i'}} = y_1czc{(zc)}^\omega y_2$. On the other hand, $zc$ can be
factorized as $zc = z_1z_2c$ such that the label $z_2cz_1$ loops on
$p$. Then the factorization of ${_0x_{i'}}$ can be now written
${_0x_{i'}} = y_1czcz_1{(z_2cz_1)}^\omega y_2$ where the first cut associated
to state $p$ is the cut $y_1czcz_1 \cdot {(z_2cz_1)}^\omega y_2$. Assume that
this cut is not secondary. Then there is a secondary cut of the form
$y_1czcz_1 {(z_2cz_1)}^n \cdot {(z_2cz_1)}^\omega y_2$ for some $n \ge 1$.
But, by Lemmas~\ref{lem:factproduct} and~\ref{lem:factomega}, either there
is no secondary cut within $z^{\prime \omega} $ or the first one occurs
before the cut $y_1c_1z_2cz_1 \cdot {(z_2cz_1)}^\omega y_2$ and this is a
contradiction.

So, we have proved that the first cut of $_0x_{i'}$ mapped to $p$ is a
secondary cut. We now show that each time the run reaches again $p$ in the
loop, the corresponding cut is a secondary cut as well. Let $z$ be the
label of the loop including~$p$. It follows that $x$ can be factorized as
$x = y_1z^\omega y_2$. Due to the duplication, the state~$p$ does not occur
in the run on the prefix $y_1z$. Let
$z = v_1^{\beta_1} \cdots v_k^{\beta_k}$ be the prime factorization
of~$z$. By Lemma~\ref{lem:factomega}, the prime factorization of~$z^\omega$
is $z^\omega = v_1^{\beta_1} \cdots v_j^{\beta_j} v^\gamma$ where a power
of $v$ is a conjugate of $z$. Note that, in this factorization, the
secondary cuts of the form $y_1zz' \cdot z''y_2$ with $z'z''= z^\omega $
occur in cuts of $v^\omega$ of the form
$v_1^{\beta_1} \cdots v_i^{\beta_i} v^{\gamma_1} \cdot v^{\gamma_2}$ where
$\gamma_1 + \gamma_2 = \gamma$. By Lemma~\ref{lem:factproduct}, in the
prime factorization of~${_0x_{i'}}$, either $v$ is a prime factor or
$v^\omega$ is a factor of a prime factor.  In the latter case, no cut of
the form $y_1zz' \cdot z''y_2$ would be secondary. This is impossible
because the cut responsible of the addition of $p$ in $Q_S$ is such a cut.
So, the prime word $v$ is a factor of the prime factorization of
${_0x_{i'}}$ and, each cut mapped to~$p$ is a cut of $z^\omega$ of the form
$v_1^{\beta_1} \cdots v_i^{\beta_i} v^{\gamma_1} \cdot v^{\gamma_2}$ which
is indeed a secondary cut. Hence, the secondary cuts exactly correspond to
the cuts mapped to a state in $Q_S$. Invariant~1 is satisfied.

Invariants 5 and 6 are trivial because the situation is the same than when
the algorithm started: the history is reduced to a single pair and the word
${_{(i',i')}x _{(k,k')}}$ is a single letter.

\subsection{Correctness and complexity}\label{sec:algoproof}

We call \emph{step} of the algorithm one execution the main while loop.
Each step falls in one three cases listed above.  We first show that the
algorithm terminates in at most~$n^4$ steps where $n$ is the number of
states of $\mathcal{A}_{\tau(e)}$.  In cases 1 and~2, the variable~$i$
remains unchanged and, in case~3, this variable is always updated to a
greater values due to Invariant~6.  This shows that case~3 cannot happen
more than $n$ times.  Between two occurrences of case~3, variable~$j$
remains unchanged in case~1 and is updated to a greater value in case~2 due
to Lemma~\ref{lem:jforward}.  This shows that case~2 cannot happen more
that $n$ times between two consecutive occurrences of case~3.  Each step
using case~1 adds a new pair to the history.  Therefore, case~1 cannot
happen more that $n^2$ times consecutively.  Putting everything together
yields the result.

We now show that the number of steps of the algorithm is at most $n^3$.  To
prove this, it is enough to remark the following fact.  Each value of the
variable~$j$ is greater than the current value of~$i$ and less that the
next value of~$i$.  This shows that the numbers of values of the pair
$(i,j)$ is less than~$n$.  Therefore, the total number of steps using cases
2 and~3 is at most $n$.  This yields that the number of steps is at most
$n^3$.

We now show that the algorithm is correct.  This means that a cut is a main
(resp., secondary) cut if and only if it is mapped to a state in~$Q_M$
(resp., $Q_S$) in the run of~$x$ in~$\mathcal{A}_{\tau(e)}$. Invariant~1
ensures that the prime factorization of~$x$ is
$x = u_1^{\alpha_1} \cdots u_k^{\alpha_k}$ where each main cut is mapped to
a state in~$Q_M$.  Invariant~6 ensures that states in~$Q_M$ never occur in
a loop of~$\mathcal{A}_{\tau(e)}$.  This implies that each of them occurs
once in the run.  This proves that other cut is mapped to a state in~$Q_M$.

We now prove that the cuts mapped to a state in~$Q_S$ are exactly the
secondary cuts.  By invariant~1, secondary cuts are mapped to states
in~$Q_S$.  The converse remains to be proved.  The proof is carried out in
two steps.  First we show that, given $q \in Q_S$, the first cut mapped
to~$q$ is secondary.  Second we show that all cuts mapped to~$q$ are
secondary.  If $q$ is not in a loop, $q$ occurs only once in the run and
the result is obvious.  Now assume that $q$ belongs to a loop whose entry
state is~$p$.

Now consider a cut $x = x_1x_2$ mapped to~$q$.  Let $z$ be the label of the
run from $p$ to~$p$.  This word~$z$ can be factorized as $z = z_1z_2$ where
$z_2z_1$ is the label of the run from~$q$ to~$q$.  Then $x$ can be
factorized as $x = y_1{(z_1z_2)}^\omega y_2$.  Due to the duplication, the
first occurrence of~$q$ in the run is just before the second occurrence
of~$z_2$.  Consider then the corresponding cut
$y_1z_1z_2z_1\cdot {(z_2z_1)}^\omega y_2$ and suppose that it is not
secondary.  Then the secondary cut $x = x_1x_2$ can be written
$x_1 = y_1z_1{(z_2z_1)}^n$ for $n \ge 2$ and $x_2 = {(z_2z_1)}^\omega y_2$.
Hence, we have a secondary cut of $x$ lying after the second occurrence
of~$z$ in $x = y_1z^\omega y_2$.  By Corollaries~\ref{cor:cutsproduct}
and~\ref{cor:cutsomega}, this not possible.  So, we have proved that the
first time the run of $\mathcal{A}_{\tau(e)}$ reaches state $q$ in the
loop, it corresponds to a position in $x$ which is indeed a secondary
cut. We now show that each time the run reaches again $q$ in the loop, the
corresponding position is a secondary cut as well. The label of the loop
including $q$ is $z =z_1z_2$ and we know that $x$ can be factorized as
$x = y_1z^\omega y_2$. The label $z$ has a prime factorization
$z = v_1^{\beta_1} \cdots v_k^{\beta_k}$. By Lemma~\ref{lem:factomega},
$z^\omega$ has the prime factorization
$z = v_1^{\beta_1} \cdots v_i^{\beta_i} v^\beta$ where a power of $v$ is a
conjugate of $z$. Note that, in this factorization, the secondary cuts that
are not within the first occurrence of~$z$ exactly occur in cuts of the
form $v_1^{\beta_1} \cdots v_i^{\beta_i} v^n \cdot v^\beta$. By
Lemma~\ref{lem:factproduct}, we know that in the prime factorization of
$x$, either $v$ is a prime factor or $v^\omega$ is a factor of a largest
prime factor. If the second case was to happen, then no secondary cut would
occur within $z^\omega$. This is impossible because we already have such a
cut in the second occurrence of $z$.  So, the prime word~$v$ is a factor of
the prime factorization of $x$ and, each time the run
of~$\mathcal{A}_{\tau(e)}$ reaches state~$q$, it is associated to a cut of
the form $v_1^{\beta_1} \cdots v_i^{\beta_i} v^n \cdot v^\beta$ which is
indeed a secondary cut. Hence, the secondary cuts exactly correspond to the
cuts mapped to a state in~$Q_S$.
\end{document}